\documentclass[%
 reprint,
 superscriptaddress,
 amsmath,amssymb,
 aps,
floatfix,
longbibliography]{revtex4-1}

\usepackage{graphicx}
\usepackage{bm}

\usepackage{xcolor}
\usepackage[colorlinks=true,citecolor=blue,linkcolor=magenta]{hyperref}

\usepackage{verbatim}

\newcommand{\nn}{\nonumber}

\newcommand{\sgn}{\operatorname{sgn}}

\newcommand{\rar}{\rightarrow}

\newcommand{\ben}{\begin{eqnarray}}
\newcommand{\een}{\end{eqnarray}}

\newcommand{\be}{\begin{equation}}
\newcommand{\ee}{\end{equation}}

\newcommand{\Arctan}{\text{Arctan}}

\begin{document}
\title{Perturbative approach to time-dependent quantum systems and applications to one-crossing multistate Landau-Zener models% with a single level-crossing point
}

\author{Rongyu Hu}
\affiliation{School of Physics and Electronics, Hunan University, Changsha 410082, China}
\author{Chen Sun}
\email{chensun@hnu.edu.cn}
\affiliation{School of Physics and Electronics, Hunan University, Changsha 410082, China}

\begin{abstract}
We formulate a perturbative approach for studying a class of multi-level time-dependent quantum systems with constant off-diagonal couplings  and diabatic energies being odd functions of time. Applying this approach to a general multistate Landau-Zener (MLZ) model with all diabatic levels crossing at one point (named the one-crossing MLZ model), we derive analytical formulas of all its transition probabilities up to $4$th order in the couplings. %These Our work predicts tractable analytical solutions  which are asymptotically exact in the diabatic limit of the multistate Landau-Zener model with a single level-crossing point,  at the same point in the diabatic limit ,   especially those ... not exactly solvable. this general type of models in the diabatic limit.
For one-crossing MLZ models it is difficult to obtain such analytical results by other kinds of approximation methods; %, so they are valuable whenever exact solutions are absent;
thus, these perturbative results can serve as reliable benchmarks for future studies of any one-crossing MLZ models that have not been exactly solved.
\end{abstract}

\maketitle

\section{Introduction}

%Exact solvability has always been a for
Finding exact analytical solutions to %Schr\"{o}dinger equations
quantum systems with a time-dependent Hamiltonian is generally regarded as a difficult task. %Unlike a time-independent problem, the knowledge of  instantaneous eigenvalues and eigenstates in analytical forms is still insufficient to exactly solve the evolution problem (the fact that Hamiltonians at different times generally do not commute prevents a direct integration to obtain the evolution matrix in an analytical form).
This remains true for a Hamiltonian with linear time-dependence --- perhaps the simplest form of time-dependence. Such linearly time-dependent quantum models are known as the multistate Landau-Zener (MLZ) models since they are generalizations of the famous Landau-Zener (LZ) model \cite{landau,zener,majorana,stuckelberg} to systems with numbers of levels larger than $2$. Although the LZ model can be exactly solved by different methods \cite{Shevchenko-2010,Shevchenko-2023} (meaning that all its transition probabilities for an evolution from $t=-\infty$ to $t=\infty$ can be expressed in analytical forms), exact solutions of MLZ models have been found only for models with special structures of their Hamiltonians \cite{Wei-1963,DO,Hioe-1987,bow-tie,Rau-1998,GBT-Demkov-2000,GBT-Demkov-2001,%chain-2002,4-state-2002,
Rau-2003,Rau-2005,Rau-2005-2,Vasilev-2007,chain-2013,4-state-2015,6-state-2015,DTCM-2016,DTCM-2016-2,
HC-2017,cross-2017,commute,Patra-2015,quest-2017,large-class,Yuzbashyan-2018,DSL-2019,MTLZ,parallel-2020,quadratic-2021,Malla-2023%,nogo-2022
}. Typical methods to exactly solve LZ or MLZ models include usages of special functions \cite{zener}, Laplace transformations and contour integrations  \cite{majorana,DO,bow-tie,GBT-Demkov-2001}, analytical constraints \cite{HC-2017,cross-2017}, the unitary integration method \cite{Wei-1963,Rau-1998,Rau-2003,Rau-2005,Rau-2005-2}, and the recently developed usage of integrability \cite{commute,Patra-2015,quest-2017,large-class,Yuzbashyan-2018,DSL-2019,MTLZ,parallel-2020,quadratic-2021,Malla-2023}. Besides, for MLZ models not exactly solvable, exact results for some of their transition probabilities may still be obtained \cite{B-E-1993,nogo-2004,Shytov-2004,Usuki-1997,Wubs-2006,Saito-2007}.

%Despite the progresses on exact-solvable time-dependent models, existence of exact solutions are rather uncommon in the sense that they rely on special choices of the Hamiltonian's parameters...
When exact solutions are absent, one naturally turn to approximation methods or numerical simulations. %... to attack time-dependent quantum models theoretically.
For MLZ models, possible approximation approaches include the independent crossing approximation, the exact WKB approach, and the Dykhne formula. The independent crossing approximation, proposed in \cite{B-E-1993}, states that when all crossings of pairs of levels are well-separated, a transition probability between two states can be approximated by adding contributions from all semiclassical paths connecting the two states, with each contribution being a product of successive pairwise transition probabilities along a path \cite{note-phase}. The exact WKB approach proves to be a powerful method to obtain approximate solutions both in the adiabatic \cite{Suzuki-2024} and in the diabatic \cite{Aoki-2002,Shimada-2020} regime. It also assumes all level crossings to be pairwise. The Dykhne formula \cite{Dykhne-1962,Davis-1976,Dykhne-2020} predicts transition probabilities in the adiabatic regime, but its application to models with more than $2$ levels is not always valid \cite{Hwang-1977}. %independent crossing approximation...
Numerical simulation of the Schr\"{o}dinger equation is also a commonly-used method for MLZ problems \cite{Ashhab-2016,Niranjan-2020}. Numerical solutions can be useful in obtaining analytical expressions of transition probabilities by fitting \cite{Hu-5-state}, and possibly by symbolic regression  algorithms \cite{La-Cava-2021}; especially, machine learning packages \cite{Udrescu-2020,Udrescu-2020-2} have been used to find analytical solutions to MLZ models recently \cite{Ashhab-2023}.

Among MLZ models there is a special class consisting of models with all diabatic levels crossing at a single point%(see Fig.~\ref{fig:diabatic-levels} for an illustration)
; we will call this class the ``{\it one-crossing MLZ model}'' for brevity. Some physically important models belong to this class. For example, a nanomagnet with spin $S$ under linearly-changing magnetic field is described by a one-crossing MLZ model with $2S+1$ levels; such a system was realized in experiments on molecular clusters \cite{Wernsdorfer-1999}. Another example is the driven Tavis-Cummings model \cite{DTCM-2016,DTCM-2016-2} --- a basic model in quantum optics --- which describes interaction between a bosonic mode and an arbitrary size spin under a linear sweep of magnetic field. Besides physical applications, one-crossing MLZ models are also of special theoretical importance, because they can be used as building blocks to construct more complex solvable systems if they themselves are solvable \cite{quest-2017}. In \cite{cross-2017}, one-crossing MLZ models were studied in detail, and some models are exactly solved by using analytical constraints on the scattering matrices and symmetry properties. Exactly solvable one-crossing MLZ models can also be generated as degenerate limits of solvable MLZ models with parallel levels \cite{GBT-Demkov-2000,DTCM-2016}. The unitary integration method \cite{Wei-1963,Rau-1998,Rau-2003,Rau-2005,Rau-2005-2} is another powerful approach to exactly solve a one-crossing MLZ model whose Hamiltonian can be expressed in terms of generators of sufficiently simple Lie algebras. Still, all these methods require strict constraints on structures of Hamiltonians, and for most one-crossing MLZ models exact solvability seem not possible. %; for example, a one-crossing $3$-state LZ model with all-to-all interactions has no known exact solutions...
As for approximation methods, the independent crossing approximation and the exact WKB method do not work for one-crossing MLZ models with more than two levels, since they assume that all crossings are pairwise. %Solution to one-crossing MLZ models may be used as building blocks to construct solutions of more complex models ......
Therefore, %simultaneous crossing of multiple levels brings additional difficulties for solving the models, and
one-crossing MLZ models require special treatment on their on.

In this paper, we obtain analytical results for one-crossing MLZ models at the diabatic limit. We first formulate a perturbative approach for treating to a general time-dependent quantum system with constant off-diagonal couplings and diabatic energies being odd functions of time. We then apply this approach to one-crossing MLZ models and derive analytical expressions of all the transition probabilities up to $4$th order in the couplings, which become asymptotically exact at the diabatic limit. These results of probabilities are general in the sense that they work for {\it any} one-crossing  MLZ models, exactly solvable or not.

Our work is expected to contribute to theoretical studies of time-dependent quantum problems in  two ways.  First, our results for one-crossing MLZ models should be valuable for any models not exactly solved --- as discussed previously, such analytical results for one-crossing MLZ models seem not achievable by other approximation methods. Second, %as recognized by the second referee...,
the general perturbative approach has its own value --- it may be applied to other multistate time-dependent quantum models with more general forms of time-dependence instead of linear time-dependence.

This paper is organized as follows. In Section II, we formulate the general perturbative approach that expresses the transition probabilities as power series of the couplings. %In Section III, we discuss properties   of the transition probabilities for a general model...
Section III is devoted to applications of the approach to one-crossing MLZ models (with technical calculation details giving in the supplemental material); in particular, transition probabilities up to $4$th order in the couplings are obtained in analytical forms (see Eqs.~\eqref{eq:Pj<k}-\eqref{eq:P24jj}), and interpretation of these results are discussed. Section IV applies the results in Section III to specific MLZ models, especially models which have not been exactly solved. Finally, in Section V we present conclusions and discussions.

\section{General perturbative approach}

We consider an $N$-state quantum system evolving under the Schr\"{o}dinger equation with a real symmetric time-dependent Hamiltonian $H(t)$ (with $\hbar$ setting to $1$):
\begin{align}\label{}
i \frac{d\psi}{dt}=H(t)\psi,  \quad H(t)= E(t) + g A,
\label{eq:Hamiltonian}
\end{align}
where $E(t)=\operatorname{diag}(\varepsilon_1(t),\varepsilon_2(t),\ldots,\varepsilon_N(t))$ is a diagonal matrix, $A$ is an off-diagonal matrix (namely, a matrix with all diagonal elements being zero) that does not depend on time, and $g$ is a real number which we will take as the parameter of perturbation expansion. The eigenvalues of the matrix $E(t)$, namely the functions $\varepsilon_j(t)$, are called {\it diabatic energies} or {\it diabatic levels}, and the eigenstates of $E(t)$ are called {\it diabatic states}. We will assume that every  $\varepsilon_j(t)$ is odd in time, namely, the matrix $E(t)$ is odd: $E(-t)=-E(t)$. Specifically, any one-crossing MLZ model, which we will focus on later, satisfies this property. %, which we will focus on later...
The element $A_{jk}$ characterizes the interaction strength between diabatic states $j$ and $k$; we call it the {\it coupling} between the two states.

A note on the units used throughout this paper is in order. We set $\hbar$ to $1$ and  time $t$ to be dimensionless, so energy is also dimensionless. $g A_{jk}$ as a whole has units of energy, and we take both $g$ and $A_{jk}$ to be dimensionless. Therefore, no explicit units will appear throughout the text.

Let's consider the evolution operator between symmetric times $-t$ and $t$:
\begin{align}\label{eq:St}
S(t)\equiv U(t,-t)=\mathcal{T} e^{-i\int_{-t}^t H(s) ds},
\end{align}
where $\mathcal{T} $ is the time-ordering operator. Denote the transition probability corresponding to this evolution from a diabatic state $k$ to a diabatic state $j$ as $P_{jk}(t)$ (note the order of the indices in $P_{jk}(t)$). $P_{jk}(t)$ is connected to the elements of $S(t)$ by $P_{jk}(t)=|S_{jk}(t)|^2$. In the limit $t\rar \infty$, i.e. when the evolution is from infinite past to infinite future, if the differences between any diabatic energies $\varepsilon_j(t)$ goes to infinity, %then no transition can occur such that the off-diagonal couplings can be neglected (as is the case for an MLZ model...),
then the transition probabilities should approach constants: $P_{jk}(t\rar \infty)=P_{jk}$. We are especially interested in these transition probabilities at infinite times. As discussed in the introduction, for a general model it is usually difficult to obtain analytical expressions of $P_{jk}$.

\subsection{Perturbative approach}
We now formulate a perturbative approach for the model \eqref{eq:Hamiltonian} to express the transition probabilities $P_{jk}(t)$ as power series expansions of the parameter $g$ around $g=0$, namely, expansions at the diabatic limit \cite{note-diabatic}. Our approach is a generalization of Waxman's perturbative approach on  $2$-state models \cite{Waxman-1994} to multistate systems.

The approach starts from differentiating \eqref{eq:St} over $t$:
\begin{align}\label{}
i\partial_t S(t)= H(t) S(t)+ S(t) H(-t).
\end{align}
Plugging in $H(t)=E(t)+gA$ and $H(-t)=-E(t)+gA$ (recall that $E(-t)=- E(t)$), we get
\begin{align}\label{eq:St-diff}
i\partial_t S(t)=[ E(t), S(t)]+ g\{A , S(t)\},
\end{align}
where ``$[,]$'' and ``$\{,\}$'' denote commutators and anti-commutators, respectively.
Let's define a diagonal matrix
\begin{align}\label{}
\Phi(t)=\int_{0}^tE(s)ds=\operatorname{diag}(\phi_1(t),\ldots,\phi_N(t))
\end{align}
where we denoted
\begin{align}\label{}
\phi_j(t)=\int_{0}^t ds \varepsilon_j(s)
\end{align}
with $j=1,2,\ldots,N$. Namely, $\phi_j(s)$ is a  phase  accumulated by the diabatic energy $\varepsilon_j(s)$ in the time interval $(0,t)$. Following \cite{Waxman-1994}, we perform a unitary transformation from $S(t)$ to an  auxiliary operator $W(t)$ by:
\begin{align}\label{eq:St-to-Wt}
%\color{red}
 S(t)= e^{-i\Phi(t)} W(t)  e^{i \Phi(t)}.
\end{align}
This transformation does not change the modulus of each component of the matrix, so $P_{ij}(t)=|W_{ij}(t)|^2$. Plugging \eqref{eq:St-to-Wt} into the differential equation \eqref{eq:St-diff} gives a differential equation for $W(t)$ in which the commutator term is eliminated:
\begin{align}\label{eq:diff-W}
i\partial_t W(t)= g \{ \tilde A (t) , W(t)\},
\end{align}
where for notation simplicity we defined
\begin{align}\label{eq:tlide-A}
\tilde A (t) = e^{i \Phi(t)}  A  e^{-i \Phi(t)}.
\end{align}
Note that $\tilde A (t)$ is Hermitian by construction. Its elements are given by %(recall that  $e^{i \Phi(t)} =\operatorname{diag}(e^{i\phi_1(t)},\ldots,e^{i\phi_N(t)})$)
\begin{align}
&\tilde A_{jj} (t) = 0,\nn\\
&\tilde A_{jk} (t) = e^{i (\phi_j(t)-\phi_k(t))}  A_{jk},  \quad \textrm{for } j\ne k.\label{eq:tildeAjk}
\end{align}
Integrating both sides of \eqref{eq:diff-W} with respect to $t$ and noticing that $W(0)=1$ (for simplicity, we use $1$ to denote an $N\times N$ identity matrix here and later on) then gives
\begin{align}\label{eq:Wt-integral}
  W(t)=  1-ig\int_0^t ds \{\tilde A (s), W(s)\} .
\end{align}
%(Note the similarity to the Dyson series in interaction picture...)
This integral equation \eqref{eq:Wt-integral} can be solved perturbatively in orders of $g$. %To achieve this it's more convenient to consider expansions in terms of a scalar parameter so let's rewrite $A$ as $g A$, namely, we separate a (scalar) factor $g$ out of the matrix $A$.
Let's write $W(t)$ as Taylor series expansion of $g$: %... %discussion on validity of Taylor series expansion...
\begin{align}\label{}
W(t)=\sum_{n=0}^\infty W_n(t)  g^n  .
\end{align}
Plugging it into  \eqref{eq:Wt-integral}, we get
\begin{align}\label{}
\sum_{n=0}^\infty W_n(t) g^n =  1-ig\int_0^t ds\{\tilde A (s) , \sum_{n=0}^\infty W_n(s) g^n \} .
\end{align}
On the two sides  of this equation the coefficients at each order of $g$ should equal. This gives $W_0(t)=1$ and a recursion relation
\begin{align}\label{eq:recursion}
  W_{n+1}(t)  =  -i\int_0^t ds\{ \tilde A  (s) ,  W_n(s)  \}  ,
\end{align}
where $n=0,1,2,\ldots$  Thus, $W_n(t)$ can be obtained iteratively. $W_{1}(t)$ %and  $W_{2}(t)$
reads
\begin{align}\label{}
 &   W_{1}(t)  =  %-i\int_0^tds \{\tilde A (s) ,  W_0(s)  \}  =
 -2i\int_0^t  ds \tilde A (s),
%& W_{2}(t) % =  -i\int_0^tds_1 \{\tilde A (s_1) ,  W_1(s_1)  \} \nn\\
%& =  -i\int_0^t ds_1 \{\tilde A (s_1),-2i\int_0^{s_1}  ds_2 \tilde A (s_2) \}\nn\\
 % =  -2\int_0^t ds_1  \int_0^{s_1}  ds_2 \{\tilde A (s_1), \tilde A (s_2) \},\label{eq:W2}
\end{align}
and $W_{n}(t)$ for $n\ge 2$ can be expressed in terms of multiple integrals:
\begin{align}\label{eq:Wn}
  &  W_{n}(t)  = 2 (-i)^n\int_0^tds_1 \int_0^{s_1}ds_2 \ldots  \int_0^{s_{n-1}}ds_n \nn\\
 & \{\tilde A (s_1), \{\tilde A (s_2),\ldots \{\tilde A (s_{n-1}) , \tilde A (s_n)  \}\ldots\}\}.
\end{align}
$W_{2}(t)$ can be simplified using symmetry. Since $\{\tilde A (s_1), \tilde A (s_2) \}$ is symmetric in $s_1$ and $s_2$, a change of the region of the double integral  from the triangle $0<s_2<s_1<t$ to a square $0<s_1,s_2<t$ would simply double the integral, so
\begin{align}\label{}
  &  W_{2}(t)  %= -\int_0^t ds_1  \int_0^{t}  ds_2 \{\tilde A(s_1), \tilde  A (s_2) \}\nn\\
  %&=-\{ \int_0^{t}  ds_1  \tilde A (s_1), \int_0^t ds_2 \tilde A (s_2) \}
  =-2 \left[ \int_0^{t}  ds  \tilde A (s) \right]^2.
\end{align}
Thus, $W_{1}(t)$ and $W_{2}(t)$ involve only evaluation of a single integral $\int_0^{t}  ds  \tilde A (s)$. One may wonder if the same is true for a general $ W_{n}(t)$, but it turns out starting from $ W_{3}(t)$ one would have to perform multiple integrals.  $ W_{3}(t)$ reads:
\begin{align}\label{}
  &  W_{3}(t)  =  2i\int_0^t ds_1 \int_0^{s_1} ds_2  \int_0^{s_2}  ds_3\{ \tilde A  (s_1) , \{\tilde A (s_2), \tilde A (s_3) \}  \} .
\end{align}
Since $\{ \tilde A  (s_1) ,  \{\tilde A (s_2), \tilde A (s_3) \}  \}$ is not completely symmetric in $s_1$, $s_2$ and $s_3$ (the terms  $\tilde A  (s_2)  \tilde A (s_1) \tilde A (s_3)$ and $\tilde A  (s_3)  \tilde  A (s_1) \tilde A (s_2)$ are missing), the above replacement of integral region cannot be performed. If we add and subtract the two missing terms, %and the multiple integral cannot be separate to into single $\int_0^{t}  ds  \tilde A (s)$ as for $W_2(t)$.
$W_{3}(t)$ is rewritten into a more symmetric form:
\begin{align}\label{}
  &  W_{3}(t)  = % &-\int_0^t ds_1 \int_0^{s_1} ds_2  \int_0^{s_2}  ds_3 [ \tilde A  (s_2) \tilde A (s_1) \tilde A (s_3)+\tilde A  (s_3)  \tilde  A (s_1) \tilde A (s_2) ]\}\nn\\
 % &=-2i[[\int_0^t ds_1  \tilde A  (s_1) ]^3 \nn\\
% 2i[\int_0^t ds_1 \int_0^{t} ds_2  \int_0^{t}  ds_3 \tilde A  (s_1)    \tilde A (s_2)  \tilde A (s_3) \nn\\
 % &-\int_0^t ds_1 \int_0^{s_1} ds_2  \int_0^{s_1}  ds_3   \tilde A  (s_2) \tilde A (s_1) \tilde A (s_3)] \nn\\
  2i \left[\int_0^t ds \tilde A  (s) \right]^3 \nn\\
  &-2i\int_0^t ds  \left[\int_0^{s } ds_1 \tilde A  (s_1)\right] \tilde A (s) \left[\int_0^{s } ds_1 \tilde A  (s_1)\right].
\end{align}
The first term arises from the completely symmetric combination and involve only $\int_0^{t}  ds  \tilde A (s)$, whereas the second term involves a multiple integral that cannot be simplified. $W_n(t)$ for $n> 3$ involve  multiple integrals  in more complicated forms. %For specific $E(t)$ and $A$, if the integrals in $W_n(t)$ up to some $n$ can be performed analytically, then one will get analytical expressions of $W(t)$ up to that order in $g$...
%Thus, it seems that for general $E(t)$ and $A$, complexity of $W_n(t)$ for $n\ge 3$ increases constantly with $n$, making analytical analysis more and more difficult. %(For specific $E(t)$ and $A$, one could still perform  numerical calculations and obtain $W_n(t)$ iteratively...)

%However, for $N=2$, we are going to show that the situation strongly simplifies...

\begin{comment}
For $W_3(t)$, since
\begin{align}\label{}
&W_{3}(t)= -i\int_0^t ds\{ \tilde A  (s) ,  W_2(s)  \} \nn\\
&=2i\int_0^t ds_1\{ \tilde A  (s_1) , [\int_0^{s}  ds_2  \tilde A (s_2)]^2 \} \nn\\
&=2i\int_0^t ds_1 \{\tilde A  (s_1)   [\int_0^{s}  ds_2  \tilde A (s_2)]^2 +h.c.\},
\end{align}
we have
\begin{align}\label{}
&W_{3,jk}(t)= -i\int_0^t ds\{ \tilde A  (s) ,  W_2(s)  \} \nn\\
&=2i\int_0^t ds_1\{ \tilde A  (s_1) , [\int_0^{s}  ds_2  \tilde A (s_2)]^2 \} \nn\\
&=-2 i \{\sum_{l,p} \int_0^t ds_1   \tilde A_{jl}(s_1) [\int_0^{s_1} ds_2   \tilde A_{lp}(s_2)][  \int_0^{s_1} ds_3 \tilde  A_{pk}(s_3)] \nn\\
&+(j\leftrightarrow k)^*\},
\end{align}
where $(j\leftrightarrow k)^*$ means to switch $j$ and $k$ and take complex conjugate in the previous  expression.
\end{comment}

Let's denote  the matrix elements of $W_{n}(t)$ as  $W_{n,jk}(t)$. In terms of $W_{n,jk}(t)$, the transition probabilities $P_{jk}(t)$ can be expressed as series expansions of $g$:
\begin{align}\label{eq:PinW}
&P_{jk}(t)=\left|\sum_{n=0}^\infty  W_{n,jk}(t)g^n\right|^2
%&= |W_{0,jk}(t)|^2+ [W_{0,jk}(t) W^*_{1,jk}(t)+W_{1,jk}(t) W^*_{0,jk}(t)] g\nn\\
%&+ [W_{0,jk} (t) W^*_{2,jk}(t) +|W_{1,jk}(t) |^2  + W_{2,jk} (t) W^*_{0,jk}(t)]  g^2\nn\\
%&+\ldots\nn\\
 =\sum_{n=0}^\infty P_{n,jk}(t) g^n,
\end{align}
where
\begin{align}\label{eq:Pjk}
&P_{n,jk}(t)=\sum_{m=0}^n W_{m,jk}(t)  W^*_{n-m,jk}(t).
\end{align}
Thus, the expansion coefficient $P_{n,jk}(t)$ is determined by all $W_{m,jk}(t)$ with $m\le n$. Taking the $t\rar\infty$ limit, we get series expansions for probabilities at infinite times:
\begin{align}\label{}
&P_{jk} \equiv \lim_{t\rar\infty} P_{jk}(t) =\sum_{n=0}^\infty  P_{n,jk} g^n,
\end{align}
where we defined
\begin{align}\label{}
&P_{n,jk} \equiv \lim_{t\rar\infty} P_{n,jk}(t) .
\end{align}
We will denote the matrix formed by $P_{n,jk}(t)$ as $P_{n}(t)$, and its limit at $t\rar \infty$ simply by $P_{n}$. %One may only be interested in the probabilities at $t\rar \infty$; in this case, (for specific $E(t)$ and $A$) if the integrals in $W_{n}(\infty)$ up to some order $n$ can be performed analytically, then  $P_n \equiv  P_{n}(\infty) $ up to that order  can also be obtained in analytic forms... %Later we will demonstrate that for MLZ models this is true for $P_n $ at least up to $n=3$.

We end this subsection by a remark on a different perturbative approach. We started from considering an evolution operator $S(t)\equiv U(t,-t)$ between symmetric times (see Eq.~\eqref{eq:St}). One could also take a different perturbative approach by considering an evolution between nonsymmetric times, namely $U(t_2,t_1)$, and finally take $t_1\rar -\infty$ and $t_2\rar \infty$. The intermediate procedures in such a approach will be different, but we expect that it will give the same final results for the probabilities at infinite times (i.e. $P_{jk}$) as the current approach, namely,
\begin{align}\label{}
\lim_{t_1 \rar-\infty , t_2\rar\infty}  |U_{jk}(t_2,t_1)|^2=\lim_{t\rar\infty}  |U_{jk}(t,-t)|^2=P_{jk}.
\end{align}
One can understand this as follows. In a sense, the two limits correspond to the ordinary value and the Cauchy principal value of an infinite integral from $-\infty$ to $\infty$, respectively. If the two limits both exist, they must equal, since the second limit is a special case of the first limit (in other words, if they do not equal, the first limit should not exist at all).

%Connection to an approach by expansion with one time fixed...

%\subsection{Properties of expansions of the probabilities}

Before considering MLZ models, we discuss some properties of $P_n(t)$ for a general model \eqref{eq:Hamiltonian}. All properties discussed in the following two subsections hold at any $t$, so they hold especially in the limit $t\rar \infty$. They will be useful when considering MLZ models in the next section.

\subsection{Structure in terms of connectivity graphs}

We first make an interesting observation on an intuitive structure of the matrix elements of $W_n(t)$ in \eqref{eq:Wn}, and further of $P_{n,jk}(t)$ in \eqref{eq:Pjk}.

\begin{figure}[!htb]
  \scalebox{0.35}[0.35]{\includegraphics{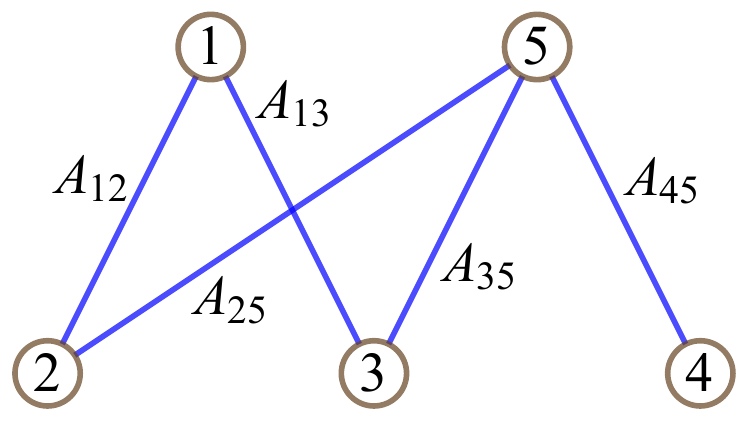}}
%\hspace{-2mm}\vspace{-4mm}%!!
\caption{The connectivity graph for a $5$-state model with non-zero couplings $A_{12}$, $A_{13}$, $A_{25}$, $A_{35}$, and $A_{45}$. }
\label{fig:connectivity-graph}
\end{figure}

For this purpose, we introduce the notation of a {\it connectivity graph}: given any $N$-state model \eqref{eq:Hamiltonian}, we associate to its Hamiltonian a graph (in the graph-theoretical sense), with each vertex labelled by $j$ corresponding to a diabatic level $j$, and each edge corresponding to a {\it non-zero} coupling $A_{jk}$ between a pair of levels $j$ and $k$. %Each coupling is labelled on its corresponding edge.
Such a graph illustrates how the levels are connected by couplings, hence the name ``connectivity graph''. For example, Fig.~\ref{fig:connectivity-graph} shows a connectivity graph for a $5$-state model.

It turns out that connectivity graphs are helpful in understanding the structure of the elements $W_{n,jk} (t)$ of the matrix $W_n(t)$ in \eqref{eq:Wn}. Despite the appearance of multiple integrals and anti-commutators, we observe that dependence of $W_{n,jk} (t)$ on the couplings is actually of a simple form. This observation follows simply from the definition of matrix products. Consider for example a product of three matrices $X$, $Y$ and $Z$, its matrix elements is given by a summation $(X Y Z)_{jk}= \sum_{l,m} X_{jl} Y_{lm} Z_{mk}$. Similarly, we can write $W_{n,jk} (t)$ as such a summation, and note that for each term in this summation there are always $n$ factors of the couplings $A_{lm}$ which do not depend on time and thus can be ``pulled out of'' the integrals (recall the definition of $\tilde A_{jk}$ in \eqref{eq:tildeAjk}). Therefore, $W_{n,jk} (t)$ must take the following form
\begin{align}\label{eq:Wnjk-path}
W_{n,jk} (t)=\sum_{l_1,l_2,\ldots,l_{n-1}}  I_{l_1 l_2\ldots l_{n-1}} A_{jl_1} A_{l_1 l_2} \ldots A_{l_{n-1} k} ,
\end{align}
where the sum is taken over all possible choices of $1 \le l_1,l_2,\ldots,l_{n-1}  \le N$, and $I_{l_1 l_2\ldots l_{n-1}}$ involves $n$-dimensional multiple integrals that depend on the diabatic energies but not on the couplings. Now note that if any factor in a certain product $ A_{jl_1} A_{l_1 l_2} \ldots A_{l_{n-1} k}$ is zero, then contribution of this product to the sum \eqref{eq:Wnjk-path} vanishes. Thus, every different term in \eqref{eq:Wnjk-path} corresponds to a different length-$n$ path from the vertex $j$ to the vertex $k$ in the connectivity graph. By identifying  each length-$n$ path in the connectivity graph, one can readily read off the structure of $W_{n,jk} (t)$. For example, in the connectivity graph in Fig.~\ref{fig:connectivity-graph}, let's identify the structure of $W_{n,12} (t) $ for $n=1,2,3$. For $n=1$, there is one length-$1$ path $1\rar2$, corresponding to a single contribution to $W_{1,12} (t) $ proportional to $A_{12}$. For $n=2$, there is no length-$2$ path, so we  know immediately that $W_{2,12} (t) =0$. For $n=3$, there are four length-$3$ paths: $1\rar2\rar1 \rar2$, $1\rar2\rar5 \rar2$, $1\rar3\rar1 \rar2$ and $1\rar3\rar5 \rar2$ (note that the paths do not need to be self-avoiding), each corresponding to a contribution to $W_{3,12} (t) $ proportional to $A_{12}^3$, $A_{12} A_{25}^2$, $A_{12} A_{13}^2$, and $A_{13} A_{35} A_{25} $, respectively (note that $A_{lm}= A_{ml}$).

The above structure of $W_{n,jk} (t)$ in turn determines the structure of $P_{n,jk} (t)$ via \eqref{eq:Pjk}. Since $P_{n,jk} (t)$ is written as a sum of $W_{m,jk}(t)  W^*_{n-m,jk}(t)$, every different pair of paths from $j$ to $k$ with lengths $m$ and $n-m$ (where $m$ ranges from $1$ to $n-1$) in the connectivity graph contributes a term in $P_{n,jk} (t)$. Equivalently, every different length-$n$ cycle that passes the vertices $j$ and $k$ corresponds to a contribution to $P_{n,jk} (t)$. For example, in the connectivity graph in Fig.~\ref{fig:connectivity-graph}, to read off the form of $P_{4,12} (t) $, note that there are 4 distinct length-$4$ cycles that passes vertices $1$ and $2$: $1\rar2\rar1 \rar2\rar1$, $1\rar2\rar5 \rar2\rar1$, $1\rar3\rar1 \rar2\rar1$ and $1\rar3\rar5 \rar2\rar1$ (again the cycles do not need to be self-avoiding), each corresponding to a contribution to $P_{4,12} (t) $:
\begin{align}\label{eq:P412-cycle}
&P_{4,12} (t)= c_1 A_{12}^4+ c_2 A_{12}^2 A_{25}^2  +c_3 A_{12}^2 A_{13}^2 \nn\\
&+  c_4 A_{12} A_{13} A_{35} A_{25}  ,
\end{align}
where $c_1$ to $c_4$ are coefficients which do not depend on any couplings $A_{lm}$.

This interpretation of $P_{n,jk} (t)$ on connectivity graphs will be useful in later sections when we deal with MLZ models. Here we mention one of its direct consequences --- if two diabatic levels $j$ and $k$ are not directly coupled (i.e. $A_{jk}=0$), then $P_{n,jk} (t)$ for $n\le 3$ all vanish, and the leading order contribution to $P_{jk} (t)$ is at least of the 4th order in $g$ (we write this briefly as $P_{jk} (t)=O(g^{4})$). More generally, if the shortest path(s) between two vertices $j$ and $k$ in the connectivity graph contains $d$ edges (namely, the ``distance'' between the two levels is $d$), then $P_{n,jk} (t)$ for $n\le 2d-1$ all vanish, and $P_{jk} (t)=O(g^{2d})$. For example, in  Fig.~\ref{fig:connectivity-graph} the distance between vertices $1$ and $5$ is $2$, so $P_{15} (t)=O(g^{4})$; the distance between vertices $1$ and $4$ is $3$, so $P_{14} (t)=O(g^{6})$.

\subsection{Symmetry properties}

Besides the interpretation in terms of connectivity graphs, $P_{n}(t)$ also have some symmetry properties. They follow directly from an interesting observation on $W_{n}(t)$. Taking Hermitian conjugate of the recursion relation \eqref{eq:recursion} and using the fact that  $\tilde A  (s)$ is Hermitian gives:
\begin{align}\label{}
    W_{n+1}^\dag (t)   =  i\int_0^t ds\{ \tilde A  (s) ,  W_n^\dag (s)\}  .
\end{align}
Since $W_0(t)=1$ is Hermitian, we obtain
\begin{align}\label{eq:Wn-property}
&   W_{n}^\dag (t)   = W_{n} (t),\quad {\textrm{for }} n \in\textrm{even} ,\\
&   W_{n}^\dag (t)   =- W_{n} (t),\quad {\textrm{for }} n \in \textrm{odd} .
\end{align}
namely, $ W_{n} (t) $ is Hermitian/anti-Hermitian for even/odd $n$.

Several properties of $P_{n}(t)$ follows. Using \eqref{eq:Pjk}, for $ P_{n}(t)$ with an even $n$,
\begin{align}\label{eq:Pn-even}
 &P_{n,kj}(t)= \sum_{m=0}^n W_{m,kj}(t)  W^*_{n-m,kj}(t) %\nn\\ &= \sum_{m=0}^n W_{m,jk}(t)^*  W_{n-m,jk}(t)
 =P_{n,jk}(t),
\end{align}
and for $ P_{n}(t)$ with an odd $n$,
\begin{align}\label{eq:Pn-odd}
&P_{n,kj}(t)= \sum_{m=0}^n W_{m,kj}(t)  W^*_{n-m,kj}(t) %\nn\\ &=- \sum_{m=0}^n W_{m,jk}(t)^*  W_{n-m,jk}(t)
 =-P_{n,jk}(t).
\end{align}
Namely, the matrix  $ P_{n} (t) $ is symmetric/anti-symmetric for even/odd $n$. Therefore, evaluation of $ P_{n,jk} (t) $ for all $j\le k$ is sufficient to determine the whole matrix $ P_{n} (t) $.

A direct consequence of \eqref{eq:Pn-odd} is that
\begin{align}\label{}
 &P_{n,jj}(t)= 0, \quad {\textrm{for }} n \in\textrm{odd} ,
\end{align}
namely, the diagonal elements of $P(t)$ can only contain terms of even orders in $g$. In other words, $ P_{jj}(t)$  must take a form:
\begin{align}\label{}
 P_{jj}(t)= 1+P_{2,jj}(t)g^2 +P_{4,jj}(t)g^4+\ldots
\end{align}

The off-diagonal elements of $P(t)$ also possess certain properties. First, since $W_0(t)=1$ one has $W_{0,jk}(t)=0$ for $j\ne k$, and it follows directly from \eqref{eq:PinW} that the leading  non-vanishing term of $P_{jk} $ must be at least of 2nd order in $g$:
\begin{align}\label{}
 P_{jk}(t)= P_{2,jk}(t)g^2 +P_{3,jk}(t)g^3 +\ldots,  \quad \textrm{for } j\ne k.
\end{align}
Second, note that $P(t)$, as a matrix of probabilities, is doubly stochastic; especially $P_{jk}(t)$ at any $g$ cannot be negative. Let's denote the leading  non-vanishing order of $P_{jk}(t)$ as $n_l$ (excluding the extremal case that $P_{jk}(t)$ is identically zero). If $n_l$ is an odd number, then at sufficiently small $g$ either $P_{n_l,jk}(t) $ or $P_{n_l,kj}(t) $ is negative, which violates the doubly stochastic property. So $n_l$ must be even --- the leading  non-vanishing term  of $P_{jk} (t)$ must be of an even order in $g$. % extreme case of $P_n{jk} (t)=0$ for any $n$.

\section{Applications to Multistate Landau-Zener model}

\begin{figure}[!htb]
\scalebox{0.45}[0.45]{\includegraphics{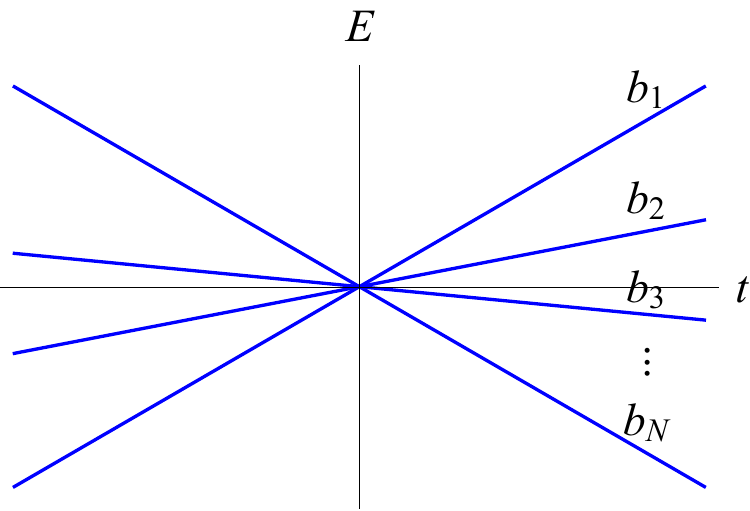}}
%\hspace{-2mm}\vspace{-4mm}%!!
\caption{Diabatic energy diagram of a one-crossing MLZ model with slopes $b_1>b_2>\ldots > b_N$.}
\label{fig:diabatic-levels}
\end{figure}

The perturbative approach described in the Section II works for a general quantum system of the form \eqref{eq:Hamiltonian}. In this paper, we focus on its application to MLZ models, namely, models whose matrix $E(t)$ is linear in $t$. Since the diabatic energies are assumed to be odd, $E(t)$ must take the form:
\begin{align}\label{eq:Et}
E(t)=\operatorname{diag}(b_1 t,b_2 t,\ldots,b_N t).
\end{align}
The parameters $b_j$ are called {\it slopes}; we take all of them to be different and assume without loss of generality that $b_1> b_2>\ldots > b_N$. %For the special case of $b_j= b_k$, we have instead $\int_0^t ds \tilde A_{jk} (s) =   A_{jk} t$. But in this case we can always diagonalize the Hamiltonian in the subspace of the degenerate slopes so $A_{jk}=0$. Thus, we can take \eqref{eq:intA} to be generally valid with the understanding that  $ A_{jk}=0$ whenever $b_j=b_k$.
This model is a one-crossing MLZ model with the crossing of levels at the point $(t,E)=(0,0)$, and its diabatic energy diagram shown in Fig.~\ref{fig:diabatic-levels}. Note that any other one-crossing MLZ model crossing at another point $(t,E)\ne (0,0)$ can be transformed to this model by a shift of $t$ and $E$ without changing the probabilities. Thus, the model \eqref{eq:Hamiltonian} with $E(t)$ given by \eqref{eq:Et} actually presents the most general form of a one-crossing MLZ model.

In this section, we will present results of transition probabilities for this general one-crossing MLZ model up to $4$th order in the couplings. We then explain these results via the connectivity graph interpretation discussed in Section IIB. We also discuss several special cases of the initial and final levels $j$ and $k$, and a specific class of model where the general results of probabilities  reduce to simpler forms, in order to provide further physical intuition of the results.

\subsection{Results of transition probabilities up to $g^4$}
By using the perturbative approach in Section IIA, in particular \eqref{eq:Wn} and \eqref{eq:Pjk}, we calculated explicitly the matrix elements of $P_n$ for this general one-crossing MLZ model up to $n=4$, namely, transition probabilities up to $4$th order in the couplings. These calculations (especially those for the $4$th order contributions) are technical and tedious so their details are put into the supplemental material, and here we present directly the results. %Note that due to the symmetry properties \eqref{eq:Pn-even} and \eqref{eq:Pn-odd}, it suffices to determine those elements $P_{n,jk}$ with $j\le k$.

%One observes that up to $g^4$, $P$ depends on the parameters $A_{jk}$ and $b_j$ only via the combinations $\lambda_{jk}\equiv A_{jk} \sqrt{\pi/|b_{jk}|}$.
%Since $A_{jj}=0$, we also define $\lambda_{jj}=0$...
Let's define the combinations
\begin{align}\label{eq:lambdajk0}
\lambda_{jk}=\sqrt{\frac{\pi}{|b_{jk}|}}A_{jk}, \quad \textrm{for } j\ne k.
\end{align}
In terms of these $\lambda_{jk}$, the probabilities $P_{jk}$ with $j<k$ read:
\begin{align}\label{eq:Pj<k}
&P_{jk}=P_{2,jk} g^2  + P_{3,jk}  g^3  + P_{4,jk}g^4   +O(g^5),
\end{align}
where
\begin{align}\label{eq:Pj<k-234}
&P_{2,jk}=2 \lambda_{jk}^2,  \nn\\
& P_{3,jk}=2 \lambda_{jk} \left(\sum_{l}^{j<l<k  }  \lambda_{jl} \lambda_{lk} -\sum_{l}^{ l<j\textrm{ or }  l>k    }  \lambda_{jl} \lambda_{lk}      \right) ,\nn\\
& P_{4,jk}= \left( \sum_{l}^{ l<j\textrm{ or }  l>k   }  \lambda_{jl} \lambda_{lk} \right)^2  + \left( \sum_{l}^{j<l<k } \lambda_{jl} \lambda_{lk}  \right)^2 \nn\\
&-  \lambda_{jk}^2\left(\sum_{l}^{l\le k}  \lambda_{jl}^2+3\sum_{l}^{l>k} \lambda_{jl}^2+\sum_{l}^{l\ge j} \lambda_{lk}^2+3\sum_{l}^{l<j} \lambda_{lk}^2 \right) \nn\\
&-2 \lambda_{jk}\left(2\sum_{l,p  }^{p<j<l<k\textrm{ or }  j<p<k<l } -\sum_{l,p  }^{ j<p<l<k } \right.\nn\\
&\left.+ \sum_{l,p  }^{j<l<k<p\textrm{, }  l<j<p<k \textrm{ or }  p<j<k<l}\right)\lambda_{jl}\lambda_{lp}\lambda_{pk}.
\end{align}
All other probabilities are then determined by symmetry properties and the doubly stochastic property. According to the symmetry properties \eqref{eq:Pn-even} and \eqref{eq:Pn-odd}, the probabilities $P_{jk}$ with $j>k$ is connected to those with $j<k$ by:
\begin{align}\label{eq:Pj>k}
&P_{jk}= P_{2,kj}g^2 -  P_{3,kj}g^3  +  P_{4,kj} g^4  +O(g^5).
\end{align}
The diagonal probabilities are given by $P_{jj}=1-\sum_{k}^{k\ne j}P_{jk}$. In particular,
\begin{align}\label{eq:Pjj}
&P_{jj}=1+ P_{2,jj}g^2+   P_{4,jj}g^4 +O(g^6),
\end{align}
where
\begin{align}\label{eq:P24jj}
&P_{2,jj}=-2\sum_{k}^{k\ne j} \lambda_{jk}^2,  \quad  P_{4,jj}= -\sum_{k}^{k\ne j}P_{4,jk}.
\end{align}
Therefore, the probabilities up to $4$th order in the couplings can be expressed as polynomials of $\lambda_{jk}$ with integer coefficients.

\subsection{Interpretation of the results via connectivity graphs}
The results \eqref{eq:Pj<k-234}, in particular the expression of the $4$th order term $P_{4,jk}$, look complicated; but we note that each term in those expressions can be understood in terms of paths on connectivity graphs discussed in Section IIB; in this subsection we discuss this interpretation. %Let's consider the terms in  \eqref{eq:Pj<k-234} in details.

First, the form of the $2$nd order contribution $P_{2,jk}$ can be directly read-off by considering length-$1$ paths from the states $j$ to $k$ (recall that $\lambda_{jk}$ defined as \eqref{eq:lambdajk0} is proportional to $A_{jk}$). Of course there is only one such length-$1$ path if $A_{jk}\ne 0$; we denote this path as $j \rar k$. This same path appears in $W_{1,jk}(t)$ and in $W^*_{1,jk}(t)$, so there is a single term in $P_{2,jk}$ which is proportional to $A_{jk}^2$. Detailed calculations in the supplemental material show that the coefficients in front of $A_{jk}^2$ is $2\pi/|b_{jk}|$, which, upon defining $\lambda_{jk}$, then leads to the expression $2\lambda_{jk}^2$ for $P_{2,jk}$ in   \eqref{eq:Pj<k-234}.

Second, let's consider the $3$rd order contribution $P_{3,jk}$. Now we should identify one length-$1$ path and one length-$2$ path from the states $j$ to $k$. The length-$1$ paths, like before, have only one possibility: $j\rar k$. For length-$2$ paths, there can be $j\rar l \rar k$ where $l$ can take any index except $j$ and $k$. Thus, the result would be a sum of the form $P_{3,jk}=2 A_{jk}  \sum_{l} A_{jl} A_{lk} c_l$, where $c_l$ are coefficients independent of the couplings. Determining these coefficients by calculations, one obtains the result of $P_{3,jk}$ in  \eqref{eq:Pj<k-234}. The coefficients for contributions from $j<l <k$ and from $l<j$ or $l>k$ turn out to be differ by a sign, so the expression of $P_{3,jk}$ in  \eqref{eq:Pj<k-234} contains two sums.

Finally, the complicated-look expression of the $4$th order contribution $P_{4,jk}$ can be understood in the same manner. Possible contributions to $P_{4,jk}$ can arise in two ways: two length-$2$ paths, or a length-$1$ path and a length-$3$ path. Possibilities of length-$1$ and length-$2$ paths are discussed before. A length-$3$ path $j\rar l\rar p\rar k $ should involve double summation of two indices $l$ and $p$. We identify that the first line of $P_{4,jk}$ in \eqref{eq:Pj<k-234} corresponds to contributions from two length-$2$ paths $j\rar l\rar k$ to $P_{4,jk}$, whereas the last three lines of $P_{4,jk}$ in \eqref{eq:Pj<k-234} correspond to contributions from a length-$1$ path $j \rar k $ and a length-$3$ path $j\rar l\rar p\rar k$ to $P_{4,jk}$. Specifically, the second line of $P_{4,jk}$  corresponds to contributions where the length-$3$ path is self-intersecting, i.e. when $l=k$ or  $p= j$. It turns out by detailed calculations that self-intersecting and self-avoided paths have coefficients of different forms, and the order of $j, l, p, k$ also influences the coefficients, leading to the various inequality conditions of the summations.

\subsection{Transition probabilities between special levels}

Now let's consider transitions between special levels where the result  \eqref{eq:Pj<k-234} reduces to simpler forms.

Case 1: A transition between the two levels with extremal slopes, i.e. with $j=1$ and $k=N$. In this case, any other level $l$ satisfies $j<l<k$, so certain sums in \eqref{eq:Pj<k-234} vanish, leading to
\begin{align}\label{eq:Pj<k-234-extreme}
&P_{2,1N}=2 \lambda_{1N}^2,  \nn\\
& P_{3,1N}=2 \lambda_{1N}  \sum_{l}  \lambda_{1l} \lambda_{lN}    ,\nn\\
& P_{4,1N}=\left( \sum_{l} \lambda_{1l} \lambda_{lN}  \right)^2 -  \lambda_{1N}^2\left(\sum_{l}  \lambda_{1l}^2+\sum_{l} \lambda_{lN}^2\right) \nn\\
&+2 \lambda_{1N}\sum_{l,p  }^{ p<l } \lambda_{1l}\lambda_{lp}\lambda_{pN},
\end{align}
where the sums with no specified ranges should be understood as over all possible $l$.

Case 2: transitions between adjacent levels, i.e. with $k=j+1$. Eq.~\eqref{eq:Pj<k-234} now reduces to:
\begin{align}\label{eq:Pj<k-234-adjacent}
&P_{2,jk}=2 \lambda_{jk}^2,  \nn\\
& P_{3,jk}=-2\lambda_{jk}\sum_{l}  \lambda_{jl} \lambda_{lk}   ,\nn\\
& P_{4,jk}= \left( \sum_{l}   \lambda_{jl} \lambda_{lk} \right)^2 -2 \lambda_{jk}   \sum_{l,p  }^{  p<j<k<l} \lambda_{jl}\lambda_{lp}\lambda_{pk} %-  \lambda_{jk}^2\left(3\sum_{l}^{l>k} \lambda_{jl}^2 +3\sum_{l}^{l<j} \lambda_{lk}^2 +2 \lambda_{jk}^2\right)
\nn\\
&-  \lambda_{jk}^2\left(\sum_{l}^{l\le k}  \lambda_{jl}^2+3\sum_{l}^{l>k} \lambda_{jl}^2+\sum_{l}^{l\ge j} \lambda_{lk}^2+3\sum_{l}^{l<j} \lambda_{lk}^2 \right),
\end{align}
where again the sums with no specified ranges are over all possible $l$, and $k$ should be understood as $j+1$.

%These two special cases...

%, we also follow your suggestion on considering specific limits and cases. In the revised text, in Section III we added discussions on transition probabilities between special levels (e.g. between two adjacent levels, from the highest to the lowest levels), and discussions on a specific class of one-crossing MLZ models where the expressions can be reduced to simple forms --- the chain models.

\subsection{A specific class --- the chain model}
To provide more insight as well as illustrate how the connectivity graph interpretation described in Section IIIB works in more specific examples, in this subsection we consider in details a specific class of one-crossing MLZ models called the chain model, where the perturbative results of probabilities reduce to very simple forms.

\begin{figure}[!htb]
  \scalebox{0.45}[0.45]{\includegraphics{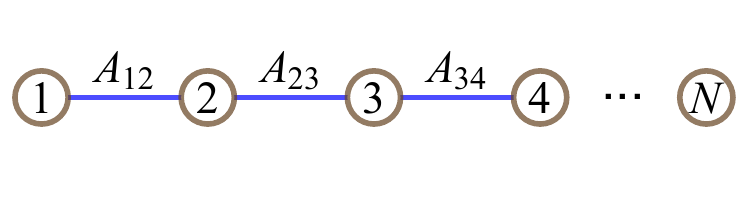}}
%\hspace{-2mm}\vspace{-4mm}%!!
\caption{The connectivity graph for an $N$-state chain model associated to the Hamiltonian \eqref{eq:hal-chain}. }
\label{fig:connectivity-graph-chain}
\end{figure}

This model has a coupling matrix $A$ with all couplings being zero except between adjacent levels: $A_{j k}= 0$ $\forall$ $|j-k|>1$. Its Hamiltonian is:
\begin{align}\label{eq:hal-chain}
&H=  \left( \begin{array}{cccccc}
b_1 t & gA_{12}   &  0  &  \cdots    & 0   & 0 \\
gA_{12} & b_2 t &  gA_{23}  &  \cdots  &   0  & 0 \\
0 &  gA_{23} & b_3 t &    \cdots  &  0  & 0 \\
\vdots &  \vdots &  \vdots  &  \ddots   & \vdots  & \vdots  \\
0&  0 &  0  &  \cdots    & b_{N-1}t  & gA_{N-1,N} \\
0 & 0  & 0  &  \cdots  & gA_{N-1,N}  & b_N t
\end{array} \right).
\end{align}
Since only couplings between adjacent levels are finite, the associated connectivity graph is a chain connecting vertices $1$, $2$, $\ldots$, $N$ successively, as shown in Fig~\ref{fig:connectivity-graph-chain}. Exact solution of a general chain model is not known. %, but for its special case --- the driven Tavis-Cummings model, exact solutions for certain tran\cite{DTCM-2016,DTCM-2016-2} as a special exactly solvable case...
Below we write out perturbative results for this chain model according to Eq. \eqref{eq:Pj<k-234} and with the help of the connectivity graph interpretation.

We first consider transitions between adjacent levels. For a transition from level $k=2$ to level $j=1$, we see that there are one length-$1$ path $1\rar 2$, no length-$2$ paths, and two length-$3$ paths $1\rar 2\rar1\rar2$ and  $1\rar 2\rar3\rar2$. The transition probability is then
\begin{align}\label{}
&P_{12}=P_{2,12} g^2  + P_{4,12}g^4   +O(g^5),
\end{align}
where
\begin{align}\label{}
&P_{2,12}=2 \lambda_{12}^2,  \nn\\
& P_{4,12}=   -  \lambda_{12}^2\left( 2\lambda_{12}^2+  \lambda_{23}^2   \right).
\end{align}
(Note that although the forms of products of $\lambda$'s can be read off on the connectivity graph, we still need Eq. \eqref{eq:Pj<k-234} to determine the coefficients in front of those products.) For a transition from level $k=j+1$ to level $j$ with $j>1$ and $j+1<N$, we see that there are one length-$1$ path $j\rar j+1$, no length-$2$ paths, and three length-$3$ paths $j\rar j+1\rar j\rar j+1$,  $j\rar j-1\rar j\rar j+1$, and $j\rar j+1\rar j+2\rar j+1$. Thus,
\begin{align}\label{}
&P_{j,j+1}=P_{2,j,j+1} g^2  + P_{4,j,j+1}g^4   +O(g^5),
\end{align}
where
\begin{align}\label{}
&P_{2,j,j+1}=2 \lambda_{j,j+1}^2,  \nn\\
& P_{4,j,j+1}=- \lambda_{j,j+1}^2\left( 2 \lambda_{j,j+1}^2+ \lambda_{j-1,j}^2+ \lambda_{j+1,j+2}^2 \right).
\end{align}
The transition probability $P_{N-1,N}$ from level $N$ to level $N-1$ is analogous to $P_{12}$, which we do not write out explicitly.

We then consider transitions between next-nearest levels, namely, from level $k=j+2$ to level $j$. Now there is no length-$1$ path and one length-$2$ path $j\rar j+1\rar j+2$. So the contribution up to $3$rd order all vanish (see the end of Section IIB):
\begin{align}\label{}
&P_{j,j+2}=P_{4,j,j+2}g^4   +O(g^5),
\end{align}
and the $4$th order contribution is of a simple form
\begin{align}\label{}
& P_{4,j,j+2}= \lambda_{j,j+1}^2  \lambda_{j+1,j+2}^2.
\end{align}

Finally, for a transition from level $k$ to level $j$ with $k-j>2$, according to the properties described at the end of Section IIB, the leading order of $P_{j,k}$ is at least $2(k-j)$, namely, $P_{j,k}= O(g^{2(k-j)})$. %Since our results are up to $4$th order, analytical form of its leading order contribution is not avaible here. beyond the current approach

%Therefore, for the chain model, leading order of transition probabilities between adjacent and next-nearest adjacent levels are of $2$nd and $4$th orders, respectively, which analytical our perturbative approaches give analytical results, whereas transition probabilities between more far away levels are of the order ...

These results apply to a general chain with arbitrary number of states $N$. At the end of Section IVB, we are going to compare these perturbative results to numerical results for a $6$-state chain model.

\section{Examples}

In this section, we apply the results of probabilities of one-crossing MLZ models, i.e. Eqs.~\eqref{eq:Pj<k}-\eqref{eq:P24jj}, to more specific models. We first discuss a few models which are exactly solvable. For these models a perturbative approach is actually not necessary, but their exact solutions can serve as tests of our perturbative results. We then consider several models whose exact solutions have not been found. We expect such models to be the arenas where our perturbative results can play important roles because, as discussed in the introduction, such analytical results are difficult to be obtained by other kinds of approximation methods. For example, the $3$-state model to be discussed in Section IVB1 describes the diabatic dynamics of a spin-$1$ nanomagnet under a fast linear sweep of magnetic field, which can guide experiments e.g. on molecular clusters \cite{Wernsdorfer-1999}.

Note that although we considered $g$ as the  parameter of perturbative expansion, $g$ can actually be absorbed into the couplings $A_{jk}$, after which the expansions can be viewed as on the couplings. %; equivalently, one simply sets $g=1$ in the original Hamiltonian in \eqref{eq:Hamiltonian}.
To make expressions simpler, in and only in this section we are going to absorb $g$ into the couplings. Formally, this means that every $A_{jk}$ appeared in this section should be understood as $g A_{jk}$ in other sections, and, as a result, every $\lambda_{jk}$ in this section should be understood as
\begin{align}\label{eq:lambdajk0g}
\lambda_{jk}=\sqrt{\frac{\pi}{|b_{jk}|}}g A_{jk}, \quad \textrm{for } j\ne k.
\end{align}

\subsection{Exactly solvable models}

The simplest and the most famous model belonging to the one-crossing MLZ class is of course the (two-state) LZ model. %As a warmup let's first consider the simplest example, namely, a model with $N=2$ levels --- the famous LZ model.
Its Hamiltonian is:
\begin{align}\label{}
& H= \left( \begin{array}{cc}
 b_1 t   &  A_{12}   \\
A_{12} &  b_2 t \\
\end{array} \right).
\label{eq:Hal-2}
\end{align}
%where we have absorbed the perturbation parameter $g$ into the coupling $A_{12}$.
The exact solutions of its transition probabilities are given by the LZ formula \cite{landau,zener,majorana,stuckelberg}:
\begin{align}\label{}
&P_{11}=P_{22}= %e^{-2\pi A_{12}^2/ b_{12} }=
e^{-2 \lambda_{12}^2 },\\
&P_{12}=P_{21}=1- e^{-2 \lambda_{12}^2 }.\label{eq:LZ-exact-P12}
\end{align}
Here, using \eqref{eq:Pj<k} and \eqref{eq:Pj<k-234} we find the series expansion of $P_{12}$ up to $4$th order of $A_{12}$: %In particular,  for the 4th order term there is no off-resonant contribution, the resonant term \eqref{eq:resonant} vanishes, and the completely symmetric term \eqref{eq:symmetric-2} is $-2\pi^2 A_{12}^4/ b_{12}^2$ which agrees with the exact result.
% $-2\pi^2 A_{12}^4/ b_{12}^2$. So there should be a mistake somewhere. fixed
\begin{align}\label{}
P_{12}=2  \lambda_{12}^2-2\lambda_{12}^4 +O(g^5),
\end{align}
which agrees with the Taylor expansion of the exact result \eqref{eq:LZ-exact-P12}. %Note that we still write $O(g^n)$ with the understanding that $g$ is absorbed in the coupling.

%This a confirmation of our result, but does not produce any new results.
\begin{table*}[]\label{}
\caption{Comparison of our perturbative results to exact solutions of the 3-state bow-tie model, the 4-state $\gamma$-magnet model, and the 4-state driven Tavis-Cummings model. References to works on the exact solutions are given under the model names. For the Tavis-Cummings model, the parameters are defined as $g_i=g\sqrt{i}$, $p_i=e^{-2\pi g_i^2}$, and $q_i=1-p_i$ for $i=1,2,3$.}
\smallskip
\begin{tabular}{|c|c|c|c|}
  \hline
  Model &  Hamiltonian & Exact solution & Perturbative solution \\
 \hline
  \begin{tabular}{c}
        \\
         3-state bow-tie  \\
               \cite{bow-tie}  \\
               \phantom{}
                     \end{tabular}
  & $\left(\begin{array}{ccc}
                       b_1t & A_{12} & 0 \\
                       A_{12} & b_2 t & A_{23} \\
                       0 & A_{23} &  b_3 t
                     \end{array}\right)
  $ & $\begin{array}{c}
                      P_{12}= (e^{-\lambda_{12}^2} +e^{-\lambda_{23}^2})(1-e^{-\lambda_{12}^2})  \\
                      P_{13}= (1-e^{-\lambda_{12}^2}) (1-e^{-\lambda_{23}^2}) \\
                     P_{23}= (e^{-\lambda_{12}^2} +e^{-\lambda_{23}^2})(1-e^{-\lambda_{23}^2})
                     \end{array} $
   & $\begin{array}{c}
                      P_{12}= 2 \lambda_{12}^2  -  \lambda_{12}^2( 2\lambda_{12}^2+  \lambda_{23}^2)+O(g^5) \\
                     P_{13}=  \lambda_{12}^2   \lambda_{23}^2  +O(g^5)  \\
                     P_{23}= 2 \lambda_{23}^2  -  \lambda_{23}^2( \lambda_{12}^2+  2\lambda_{23}^2)  +O(g^5)
                     \end{array} $ \\
  \hline
 \begin{tabular}{c}
                       $4$-state $\gamma$-magnet \\
               \cite{4-state-2015,DSL-2019}
                     \end{tabular}
& $\left( \begin{array}{cccc}
 b_1 t   &  A_{12}    &  A_{13} &  0\\
A_{12} &  b_2 t &   0 &  -A_{13} \\
 A_{13} & 0 & -b_2 t & A_{12} \\
0 & -A_{13} & A_{12} & -b_1 t
\end{array} \right)$  &
$\begin{array}{c}
                      P_{12}=P_{34}=  e^{-2\lambda_{13}^2} (1-e^{-2\lambda_{12}^2})  \\
                      P_{13}=  P_{24}= 1-e^{-2\lambda_{13}^2}  \\
                     P_{14}=P_{23}= 0
                     \end{array} $
& $\begin{array}{c}
                      P_{12}=   2\lambda_{12}^2 -2 \lambda_{12}^2(  \lambda_{12}^2 +2 \lambda_{13}^2) +O(g^5)\\
                      P_{13}=  2\lambda_{13}^2-2\lambda_{13}^4 +O(g^5)\\
                     P_{14}=O(g^5)
                     \end{array} $ \\
  \hline

   \begin{tabular}{c}
                       $4$-state driven\\
                       Tavis-Cummings  \\
         \cite{DTCM-2016}
                     \end{tabular}
  & $\left( \begin{array}{cccc}
 3 t   &  \sqrt 3 g_3  &  0 &  0\\
\sqrt 3 g_3 & 2 t &   2  g_2 &  0 \\
0 &  2  g_2 &   t & \sqrt 3 g_1 \\
0 & 0& \sqrt 3 g_1 & 0
\end{array} \right)$   &
$\begin{array}{c}
                      P_{12}=q_3(p_2^2+ p_2p_3 + p_3^2)  \\
                      P_{13}=  q_2q_3(p_1 + p_2 + p_3) \\
                     P_{14}= q_1 q_2 q_3
                     \end{array} $
 &$\begin{array}{c}
                      P_{12}= 18\pi g^2- 234\pi^2 g^4+O(g^5)  \\
                      P_{13}= 72 \pi^2 g^4+O(g^5) \\
                     P_{14}=  O(g^6)
                     \end{array} $\\
\hline
\end{tabular}
\end{table*}

Besides the LZ model, there are other classes of exactly solvable one-crossing MLZ models. In Table I, we compare the exact solutions of examples of three models --- the bow-tie model \cite{bow-tie}, the $\gamma$-magnet model \cite{4-state-2015,DSL-2019}, and the driven Tavis-Cummings model \cite{DTCM-2016} --- with our perturbative results. In all cases, we find agreement of our results to series expansions of the exact solutions. %, which supports correctness... of our results.

\subsection{Models not exactly solved}

Below we consider several one-crossing MLZ models which have not been exactly solved. We will write out explicitly probabilities up to $4$th order in the couplings. For each model, we perform numerical calculations at different choices of parameters to further confirm the analytical perturbative results.

\subsubsection{3-state LZ model with all-to-all couplings}
The simplest yet unsolved example would be a one-crossing $3$-state LZ model. Its Hamiltonian is:
\begin{align}\label{}
& H= \left( \begin{array}{ccc}
 b_1 t   &  A_{12}    &  A_{13}\\
A_{12} &  b_2 t &  A_{23}  \\
 A_{13} & A_{23} & b_3 t
\end{array} \right),
\label{eq:Hal-3}
\end{align}
where $b_1>b_2>b_3$. If in addition $b_2 =(b_1+b_3)/2$, this model describes an $S=1$ spin under linearly-changing magnetic field in the $z$-direction with arbitrary couplings between the three states. When any of the three couplings $A_{12}$, $A_{13}$, $A_{23}$  is zero, the model belongs to the exactly-solvable bow-tie model \cite{bow-tie} discussed previously. For the general case when all couplings are non-vanishing, exact solution  is not known. %Using \eqref{eq:Pdiag-upto3} and \eqref{eq:Poffdiag-upto3},
Here, for transition probabilities up to $4$th order in the couplings, we obtain for the off-diagonal elements:
\begin{align}\label{}
%&P_{12}= 2\pi  \frac{A_{12}^2}{b_{12}}-2\pi^{\frac{3}{2}}\frac{A_{12}}{\sqrt {b_{12}}}\frac{A_{13}A_{23}}{\sqrt {b_{13}b_{23}}}  +O(A_{jk}^4),\\
&P_{12}= 2  \lambda_{12}^2-2\lambda_{12}\lambda_{13}\lambda_{23} \label{eq:3-state-P12} \nn\\
&+ \lambda_{13}^2\lambda_{23}^2-\lambda_{12}^2(3\lambda_{13}^2+\lambda_{23}^2+2\lambda_{12}^2)+O(g^5),\\
%&P_{21}= 2\pi  \frac{A_{12}^2}{b_{12}}+2\pi^{\frac{3}{2}} \frac{A_{12}A_{13}A_{23}}{\sqrt {b_{12}b_{13}b_{23}}}  +O(A_{jk}^4),\\
&P_{13}= 2   \lambda_{13}^2+2\lambda_{12}\lambda_{13}\lambda_{23} \nn\\
&+\lambda_{12}^2\lambda_{23}^2 -\lambda_{13}^2(\lambda_{12}^2+\lambda_{23}^2 +2\lambda_{13}^2)+O(g^5),\\
%&P_{31}= 2\pi  \frac{A_{13}^2}{b_{13}}-2\pi^{\frac{3}{2}} \frac{A_{12}A_{13}A_{23}}{\sqrt {b_{12}b_{13}b_{23}}}  +O(A_{jk}^4),\\
&P_{23}= 2   \lambda_{23}^2-2\lambda_{12}\lambda_{13}\lambda_{23}  \nn\\
&+\lambda_{12}^2\lambda_{13}^2-\lambda_{23}^2( \lambda_{12}^2+3\lambda_{13}^2+2\lambda_{23}^2)+O(g^5),
%&P_{32}= 2\pi  \frac{A_{23}^2}{b_{23}}+2\pi^{\frac{3}{2}} \frac{A_{12}A_{13}A_{23}}{\sqrt {b_{12}b_{13}b_{23}}}  +O(A_{jk}^4).
\end{align}
and for the diagonal elements:
\begin{align}\label{}
&P_{11}=1-2( \lambda_{12}^2+\lambda_{13}^2 )+ 2 (\lambda_{12}^2+\lambda_{13}^2)^2+O(g^6),\label{eq:3-state-P11}\\
&P_{22}=1-2( \lambda_{12}^2+\lambda_{23}^2 )\nn\\
&+2(\lambda_{12}^2 \lambda_{13}^2+\lambda_{12}^2 \lambda_{23}^2+\lambda_{13}^2 \lambda_{23}^2+\lambda_{12}^4+\lambda_{23}^4)+O(g^6),\label{eq:3-state-P22}\\
&P_{33}=1-2( \lambda_{13}^2+\lambda_{23}^2 )+ 2 (\lambda_{13}^2+\lambda_{23}^2)^2+O(g^6).\label{eq:3-state-P33}
\end{align}
Due to the symmetric properties, $P_{21}$, $P_{31}$ and $P_{32}$ can be obtained by changing the signs of the $3$rd order terms in  $P_{12}$, $P_{13}$ and $P_{23}$, respectively. Note that $P_{12}$ and $P_{23}$ are connected to each other by switching of indices $1\leftrightarrow 3$, and so are  $P_{11}$ and $P_{33}$.

\begin{figure}[!htb]
\scalebox{0.33}[0.33]{\includegraphics{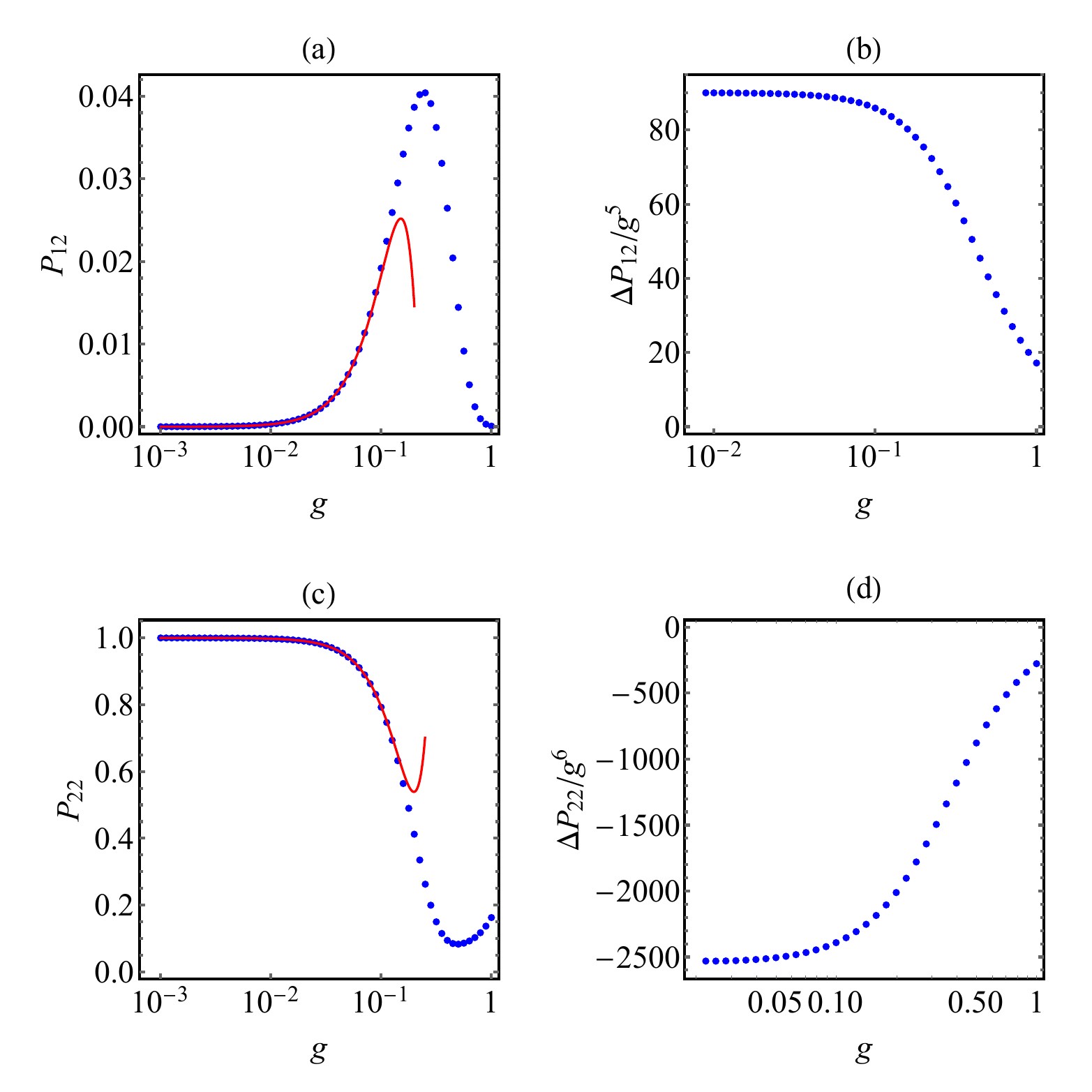}}
%\hspace{-2mm}\vspace{-4mm}%!!
\caption{Comparison of analytical series expansions and numerical results of transition probabilities of the $3$-state LZ model  \eqref{eq:Hal-3}  at parameter choices  $A_{12} = g$, $A_{13} = 1.5g$, $A_{23} = 1.8g$, $b_1 = 2$, $b_2 = 0$, and $b_3 = -1$. (a) $P_{12}$ vs. $g$ from numerical calculations (the dots) and from the series expansion expression \eqref{eq:3-state-P12} (the curve). (b) $\Delta P_{12}/g^5$ vs. $g$, where  $\Delta P_{12}$ is the difference of the numerically obtained $P_{12}$ and the series expansion expression \eqref{eq:3-state-P12}. (c) $P_{22}$ vs. $g$ from numerical calculations (the dots) and from the series expansion expression \eqref{eq:3-state-P22} (the curve). (d) $\Delta P_{12}/g^6$ vs. $g$, where  $\Delta P_{22}$ is the difference of the numerically obtained $P_{22}$ and the series expansion expression \eqref{eq:3-state-P22}.  }
\label{fig:numerics-3-state}
\end{figure}

To check the validity of these analytical expressions of the series expansion of transition probabilities, we performed numerical simulations of the Schr\"{o}dinger equation of the model \eqref{eq:Hal-3}. Comparison of the analytical series expansion  results and numerical results for two transition probabilities are shown in Fig.~\ref{fig:numerics-3-state} at one set of randomly chosen parameters. Fig.~\ref{fig:numerics-3-state}(a) plots $P_{12}$ from numerics and from the series expansion, i.e. Eq.~\eqref{eq:3-state-P12}. Their agreement at small $g$ confirms the series expansion expressions of $P_{12}$. As a more accurate check, in Fig.~\ref{fig:numerics-3-state}(b) we plot the ratio $\Delta P_{12}/g^5$, where $\Delta P_{12}$ is the numerically obtained $P_{12}$ subtracted by terms from the analytical series expansion expression \eqref{eq:3-state-P12} up to $g^4$. The fact that this ratio approaches a constant at small $g$ indicates that the difference is of the order of $g^5$, which confirms the correctness of  Eq.~\eqref{eq:3-state-P12}. (If any expansion coefficients of the numerical result up to $g^4$ were different from Eq.~\eqref{eq:3-state-P12}, that ratio would diverge at $g\rar 0$.) Fig.~\ref{fig:numerics-3-state}(c) and (d) are similar to Fig.~\ref{fig:numerics-3-state}(a) and (b) but for $P_{22}$. Note that in Fig.~\ref{fig:numerics-3-state}(d) the ratio $\Delta P_{22}/g^6$ is plotted, since  $P_{22}$, being a  diagonal element, has a zero expansion coefficient in $g^5$.

\begin{figure}[!htb]
\scalebox{0.4}[0.4]{\includegraphics{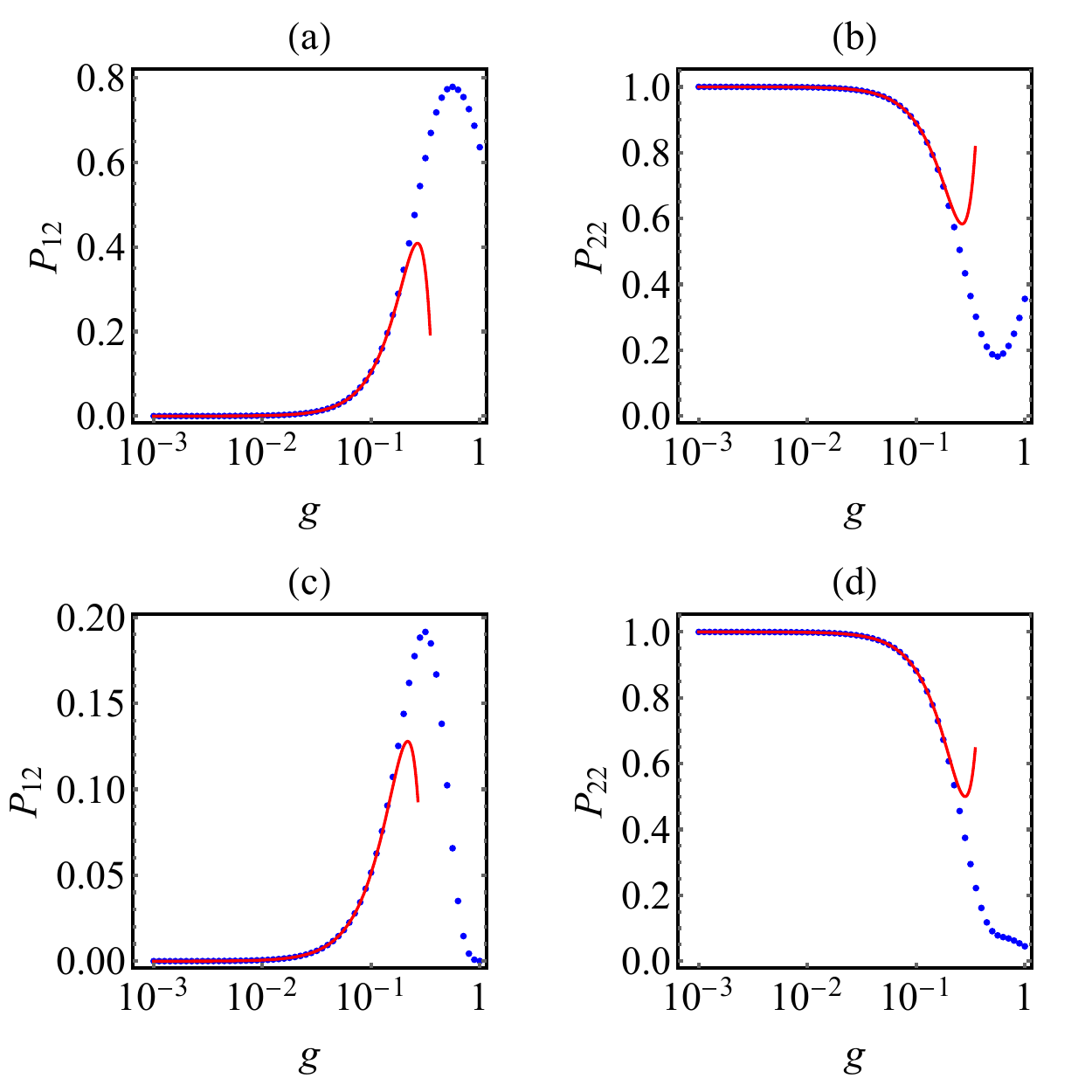}}
%\hspace{-2mm}\vspace{-4mm}%!!
\caption{Comparison of analytical series expansions (the curves) and numerical results (the dots) of transition probabilities of the $3$-state LZ model \eqref{eq:Hal-3} at two other parameter choices. (a) and (b) are at parameters $A_{12} = -1.5g$, $A_{13} = 1.1g$, $A_{23} = 0.4g$, $b_1 = 1.3$, $b_2 = 0$, and $b_3 = -1$. (a) $P_{12}$ vs. $g$ with series expansion given by \eqref{eq:3-state-P12}. (b) $P_{22}$ vs. $g$ with series expansion given by \eqref{eq:3-state-P22}. (c) and (d) are the same as (a) and (b), respectively, except that the parameters are $A_{12} = A_{13} = A_{23} = g$, $b_1 = 1$, $b_2 = 0$, and $b_3 = -1$.  }
\label{fig:numerics-3-state2}
\end{figure}

More numerics are performed at other choices of parameters: a different random choice as shown in Fig.~\ref{fig:numerics-3-state2}(a) and (b), and a ``uniform'' choice (with uniform couplings and equal slope differences between adjacent levels) as shown in Fig.~\ref{fig:numerics-3-state2}(c) and (d). They all show that the perturbative results agree with the numerical results at small $g$. Moreover, we see that the values of $g$ above which perturbative results have visible deviation from numerical results are roughly at or slightly above $g= 0.1$ (also see the footnote \cite{note-diabatic}).

Besides the numerical tests, we also compare the series expansion results with the Brundobler-Elser (BE) formula \cite{B-E-1993,nogo-2004,Shytov-2004}, which says that for any MLZ model, the probabilities to stay on the levels with extremal slopes take exact analytical forms. For a general one-crossing MLZ model (namely, a model \eqref{eq:Hamiltonian} with $E(t)$ given by \eqref{eq:Et} and $g$ set to $1$) and in terms of $\lambda_{ij}$ defined in \eqref{eq:lambdajk0g}, the BE formula predicts:
\begin{align}\label{}
&P_{11}=e^{-2\sum_{l=1}^N \lambda_{1l}^2},\\
&P_{NN}=e^{-2\sum_{l=1}^N \lambda_{lN}^2}.
\end{align}
(The LZ formula can be viewed as a special case of the BE formula at $N=2$.) Here for the $3$-state model \eqref{eq:Hal-3}, we have
\begin{align}\label{}
&P_{11}=e^{-2( \lambda_{12}^2+\lambda_{13}^2 )},\\
&P_{33}=e^{-2( \lambda_{13}^2+\lambda_{23}^2 )}.
\end{align}
Performing Taylor expansions of these two exact results, one sees that \eqref{eq:3-state-P11} and  \eqref{eq:3-state-P33} for $P_{11}$ and $P_{33}$ do agree with them.

We expect these perturbative results of the $3$-state model to be practically useful when considering the dynamics of a spin-$1$ nanomagnet under a linear sweep of magnetic field. When such a sweep is fast enough (namely in the diabatic limit) our analytical results apply, and they provide theoretical estimates of the excitation probabilities of the spin under the sweep (such analytical estimates have not been achieved by any other theoretical methods), which can further guide experiments on nanomagnets.

\subsubsection{A $4$-state LZ model}
\begin{figure}[!htb]
\scalebox{0.36}[0.36]{\includegraphics{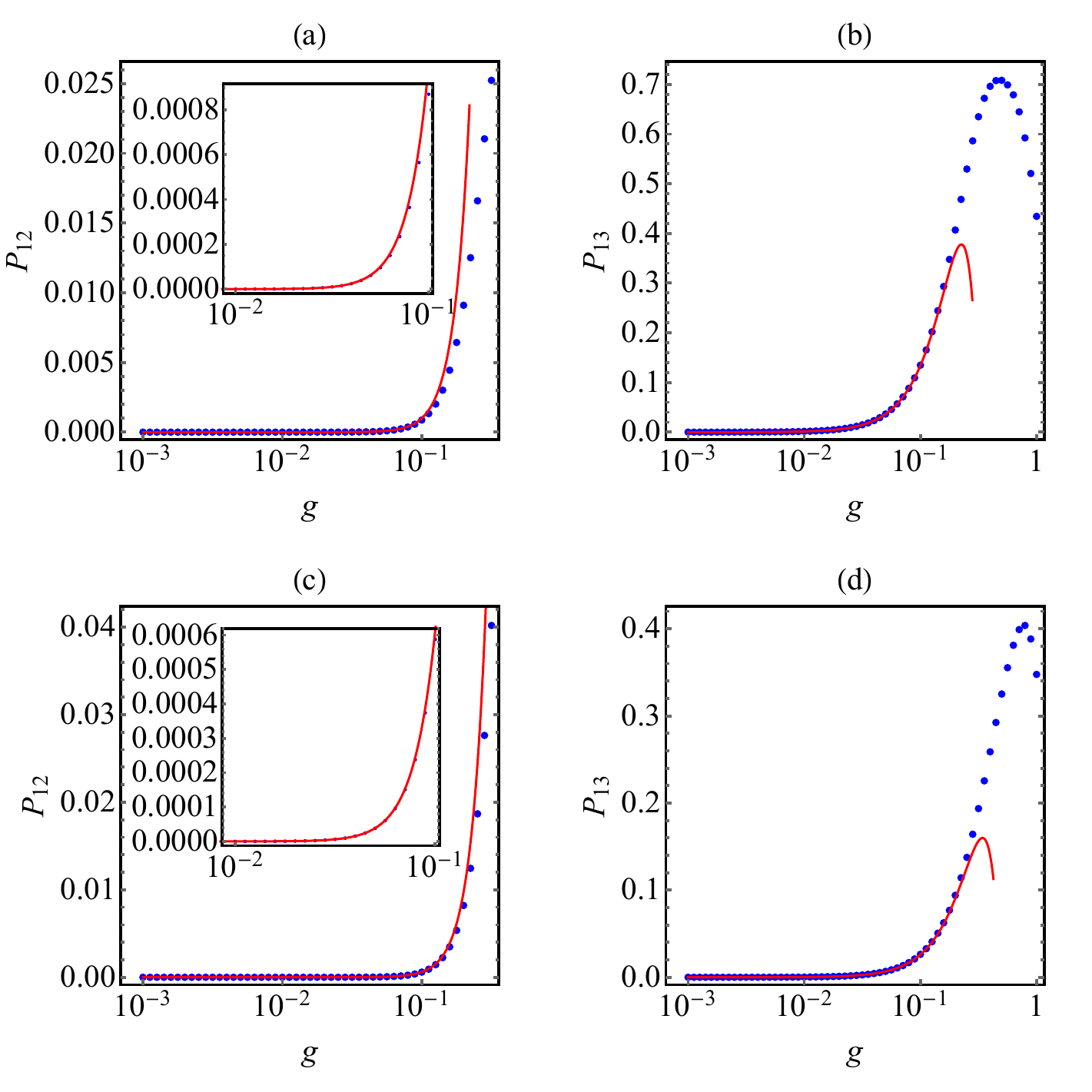}}
%\hspace{-2mm}\vspace{-4mm}%!!
\caption{Comparison of analytical series expansions (the curves) and numerical results (the dots) of transition probabilities of the $4$-state LZ model \eqref{eq:Hal-4-bipartite} at two parameter choices. (a) and (b) are at parameters $A_{13} = 3g$, $A_{23} = g$,  $A_{14} = -1.5g$,  $A_{24} = 0.4g$, $ b_1 = 3$, $b_2= 1$,  $b_3 = -0.8$, and $b_4 = -2.4$. (a) $P_{12}$ vs. $g$ with series expansion given by \eqref{eq:4-state-bipartite-P12}; the inset shows the same curve zoomed into the region $0.01<g<0.1$. (b) $P_{22}$ vs. $g$ with series expansion given by \eqref{eq:4-state-bipartite-P13}. (c) and (d) are the same as (a) and (b), respectively, except that the parameters are $A_{13} = -g$, $A_{23} = 1.2g$,  $A_{14} = 0.4g$,  $A_{24} = -0.7g$, $ b_1 = 1.5$, $b_2= 0.5$,  $b_3 = -0.8$, and $b_4 = -2$.  }
\label{fig:numerics-4-state}
\end{figure}

We next consider a one-crossing $4$-state LZ model, with a Hamiltonian:
\begin{align}\label{}
& H= \left( \begin{array}{cccc}
 b_1 t   &  0    &  A_{13} &  A_{14}\\
0 &  b_2 t &  A_{23}  & A_{24}  \\
 A_{13} & A_{23} & b_3 t &0 \\
 A_{14} & A_{24} & 0 & b_4 t
\end{array} \right),
\label{eq:Hal-4-bipartite}
\end{align}
where $b_1>b_2>b_3>b_4$. In this model two out of the six couplings are zero. We obtain:
\begin{align}
&P_{12}= (\lambda_{13}\lambda_{23} +\lambda_{14}\lambda_{24})^2+O(g^5) ,\label{eq:4-state-bipartite-P12}\\
&P_{13}= 2  \lambda_{13}^2 -\lambda_{13}^2( 3\lambda_{14}^2+ \lambda_{23}^2 +2\lambda_{13}^2)\nn\\
&-4\lambda_{13}\lambda_{14}\lambda_{23}\lambda_{24}  +O(g^5)\label{eq:4-state-bipartite-P13} ,\\
&P_{14}= 2  \lambda_{14}^2 -\lambda_{14}^2(\lambda_{12}^2+\lambda_{13}^2+ \lambda_{24}^2+2\lambda_{14}^2) \nn\\
& +2 \lambda_{13} \lambda_{14}\lambda_{23}\lambda_{24} +O(g^5) ,\\
&P_{23}= 2  \lambda_{23}^2 -\lambda_{23}^2( 3\lambda_{24}^2 +3\lambda_{13}^2 +2\lambda_{23}^2)\nn\\
&-2 \lambda_{13}\lambda_{14}\lambda_{23}\lambda_{24} +O(g^5) ,\\
&P_{11}= 1-2( \lambda_{13}^2 +\lambda_{14}^2) + 2 ( \lambda_{13}^2 + \lambda_{14}^2)^2+O(g^6) \label{eq:4-state-bipartite-P11},\\
&P_{22}=1-2( \lambda_{23}^2 +\lambda_{24}^2)\nn\\
&+\lambda_{23}^2 \left(2 \lambda_{13}^2+3 \lambda_{24}^2+2\lambda_{23}^2\right)+\lambda_{24}^2 \left(2 \lambda_{14}^2+\lambda_{23}^2\right)\nn\\
&+2 \lambda_{24}^4+4 \lambda_{13} \lambda_{14}\lambda_{23}\lambda_{24}+O(g^6).
\end{align}
Again, other $P_{jk}$ with $j\le k$ are connected to these explicitly written out ones  by switches of indices $1\leftrightarrow 4$ and $2\leftrightarrow 3$ everywhere, and $P_{jk}$ with $j>k$ can be obtained by changing the signs of the $3$rd order terms in  $P_{kj}$. Comparisons to numerics at two sets of parameters are shown in Fig.~\ref{fig:numerics-4-state}, which again show agreement at small $g$ ($g\lesssim 0.1$). Besides, $P_{11}$ in \eqref{eq:4-state-bipartite-P11} agrees with the exact result from the BE formula.

For the most general one-crossing $4$-state LZ model with all-to-all couplings, the expressions of probabilities also straightforwardly follow from Eqs.~\eqref{eq:Pj<k}-\eqref{eq:P24jj}, but are more complicated. We present results of that more general model in the supplemental material.

\subsubsection{A $5$-state LZ model}

We next consider a one-crossing $5$-state LZ model with the following Hamiltonian
\begin{align}\label{eq:hal-5-state}
&H=  \left( \begin{array}{ccccc}
-b_1 t & A_{12}   & A_{13}   &  A_{14}     & 0  \\
A_{12} & -b_2 t & 0  &  0  &  -A_{14}\\
A_{13} &  0 & 0 &    0  & -A_{13} \\
A_{14} &  0 &  0  &  b_2 t   & -A_{12}\\
0 & -A_{14}  & -A_{13}  &  -A_{12}  & b_1 t
\end{array} \right),
\end{align}
\begin{comment}
or after reordering:
\begin{align}\label{}
&H'=  \left( \begin{array}{ccccc}
 b_2 t  & -A_{12}   & 0  &  A_{14}     & 0  \\
-A_{12} & b_1 t & -A_{13}  &  0  &  -A_{14}\\
0 &  -A_{13} & 0 &    A_{13}  & 0 \\
A_{14} &  0 &  A_{13}  &  - b_1 t  & A_{12}\\
0 & -A_{14}  & 0  &  A_{12}  & -b_2 t
\end{array} \right),
\end{align}
\end{comment}
where $b_2>b_1>0$. This model is related to a $5$-site Su-Schrieffer-Heeger chain under a linear quench of its couplings, and was studied in detail in \cite{Hu-5-state}. Its exact solution has not been found, but by analytical constraint method it was shown  in \cite{Hu-5-state} that all of its transition probabilities depends only on two of them, for example, $P_{32}$ and $P_{33}$. Here we obtain for these two probabilities  (note that since the levels in the Hamiltonian \eqref{eq:hal-5-state} are not in descending order of the slopes, they should be reordered before applying Eqs.~\eqref{eq:Pj<k}-\eqref{eq:P24jj}):
\begin{align}
%&P_{31}=P'_{34}
%=2 \lambda_{jk}^2 \nn\\
%&-  \lambda_{jk}^2\left(\sum_{l}^{l<k}  \lambda_{jl}^2+3\sum_{l}^{l>k} \lambda_{jl}^2+\sum_{l}^{l>j} \lambda_{lk}^2+3\sum_{l}^{l<j} \lambda_{lk}^2 +2  \lambda_{jk}^2\right) \nn\\
%& -2 \lambda_{jk}\left(2\sum_{l,p  }^{p<j<l<k\textrm{ or }  j<p<k<l } -\sum_{l,p  }^{ j<p<l<k } \right.\nn\\
%&\left.+ \sum_{l,p  }^{j<l<k<p\textrm{, }  l<j<p<k \textrm{ or }  p<j<k<l}\right)\lambda_{jl}\lambda_{lp}\lambda_{pk}+O(A_{jk}^5)\nn\\
%&=2 \lambda_{13}^2  -  \lambda_{13}^2\left( \lambda_{13}^2+ \lambda_{12}^2+ 3 \lambda_{14}^2 +2  \lambda_{13}^2\right)+O(A_{jk}^5),\\
&P_{32}%=P'_{35}
=  \lambda_{13}^2(\lambda_{12}^2 +\lambda_{14}^2)+O(g^5) ,\label{eq:5-state-P32}\\
&P_{33}%=1-2P_{31}-2P_{32}
= 1-4 \lambda_{13}^2+ 2\lambda_{13}^2(3\lambda_{13}^2 +2\lambda_{14}^2)+O(g^6).\label{eq:5-state-P33}
\end{align}
By numerical calculations at different choices of parameters, the series expansion of $|S_{32}|$ and $S_{33}$ was found in \cite{Hu-5-state} (see Eqs.~(52) and (53) there). Taking squares of them, we find that they do agree with \eqref{eq:5-state-P32} and \eqref{eq:5-state-P33} here. Hence, the  series expansions previously found from numerics can now be derived analytically.

\subsubsection{A $6$-state chain model}
\begin{figure}[!htb]
\scalebox{0.4}[0.4]{\includegraphics{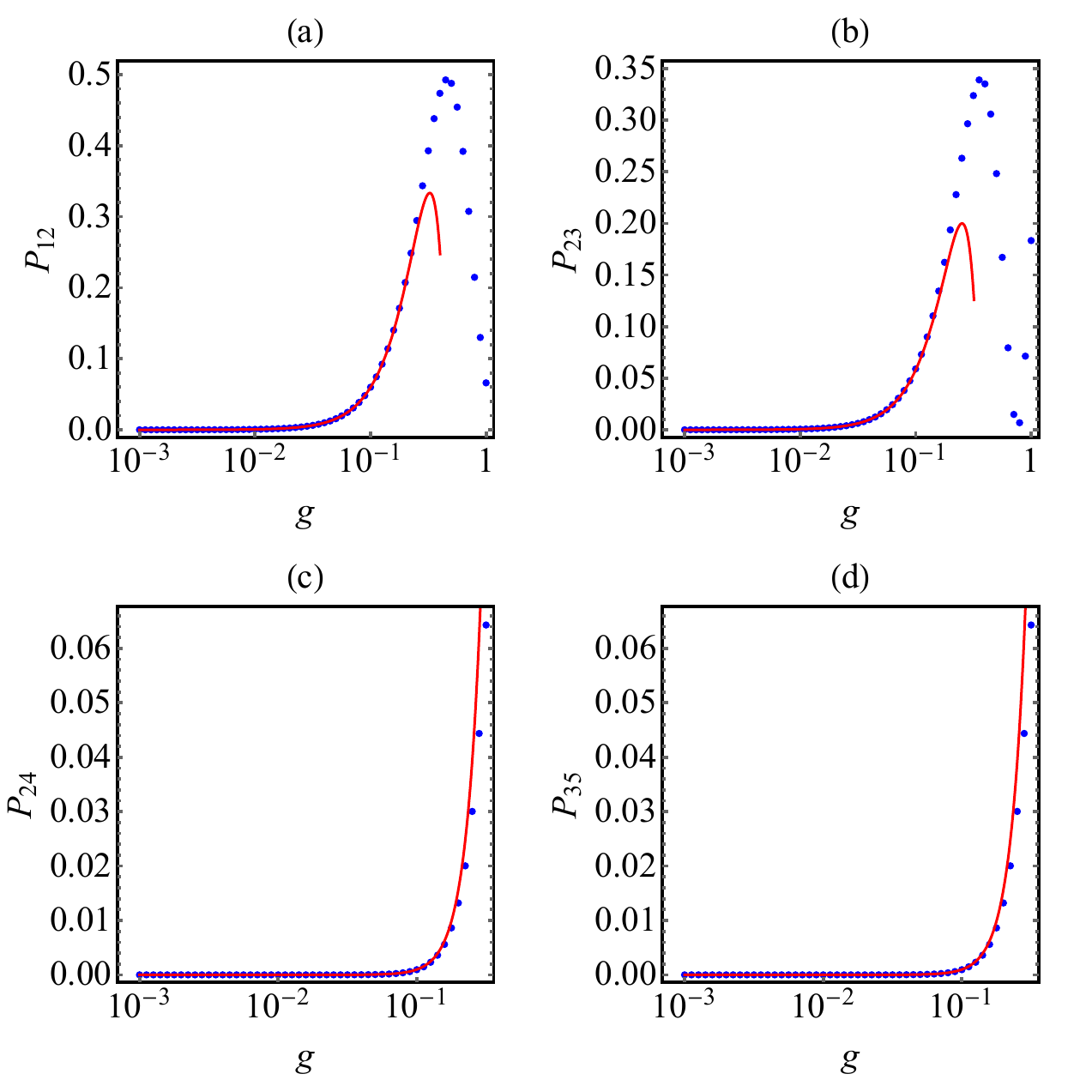}}
%\hspace{-2mm}\vspace{-4mm}%!!
\caption{Comparison of analytical series expansions (the curves) and numerical results (the dots) of transition probabilities of the $6$-state chain model \eqref{eq:hal-6-chain} at the parameter choices $A_{j,j+1} = g$ ($j=1,2,3,4,5$), $ b_1 = 2.5$, $b_2= 1.5$,  $b_3 = 0.5$, and $b_4 = -0.5$, $b_5 = -1.5$, $b_6 = -2.5$ for (a) $P_{12}$ vs. $g$; (b) $P_{23}$ vs. $g$; (c) $P_{24}$ vs. $g$; (d) $P_{35}$ vs. $g$.  }
\label{fig:numerics-6-state}
\end{figure}

As a final example, we consider a $6$-state chain model with a Hamiltonian:
\begin{align}\label{eq:hal-6-chain}
&H=  \left( \begin{array}{cccccc}
b_1 t &  A_{12}   &  0  &  0    & 0   & 0 \\
 A_{12} & b_2 t &   A_{23}  &  0  &   0  & 0 \\
0 &   A_{23} & b_3 t &    A_{34}  &  0  & 0 \\
0 &  0 &  A_{34}   & b_4 t &  A_{45}  & 0  \\
0&  0 &  0  &  A_{45}    & b_{5}t  &  A_{56} \\
0 & 0  & 0  &  0  & A_{56}  & b_6 t
\end{array} \right).
\end{align}
In Section IIID, we determined transition probabilities up to $4$th-order in couplings for a general $N$-state chain model. Here we compare these perturbative results to numerics for a $6$-state chain model with uniform couplings and equal slope differences between adjacent levels in Fig.~\ref{fig:numerics-6-state}. Again, one see agreement between the two at small $g$ ($g\lesssim 0.1$).

\section{Conclusions and Discussions}

We formulated a perturbative approach that applies to a general time-dependent quantum system with constant off-diagonal couplings and diabatic energies being odd functions of time. Using this approach for a general MLZ model with all levels crossing at a single point (one-crossing MLZ model), we derived analytical expressions of all its transition probabilities up to $4$th order in the couplings (Eqs.~\eqref{eq:Pj<k}-\eqref{eq:P24jj} in Section IIIA), which become asymptotically exact at the diabatic limit. These expressions depend on the couplings $A_{jk}$ and the slopes $b_j$  solely via the combinations $A_{jk}/\sqrt{|b_j-b_k|}$. To our best knowledge, such analytical results haven't been achieved by any other approximation methods for one-crossing MLZ models not exactly solvable. These analytical asymptotic solutions at the diabatic limit can serve as reliable references for future studies of any unsolved one-crossing MLZ models; for example, transition probabilities for any one-crossing MLZ model found in future studies by any method must have series expansions which agree with the results here.

Let us discuss possible future extensions of this work. One direction is of course trying to derive analytical expressions of higher order expansions of probabilities for one-crossing MLZ models, although we expect that calculations may be very involved (given the complexity of the performed calculations of $4$th order terms). Second, in this work we focused on MLZ models, namely, models with linear time dependence. But recall that the perturbative approach can be applied to a model with a more general time-dependence of its diabatic energies as long as they are odd in time. One may thus also consider models with diabatic energies of forms $t^3$ or $\sinh t$, etc., and the series expansion of transition probabilities are expected to be expressed in terms of multiple integrals of different forms corresponding to the specific time dependencies. Third, probabilities at finite instead of infinite times may also be considered. %For example, in \cite{Waxman-2003}, Waxman considered transition probabilities at stroboscopic times of a two-state model with a sinusoidal time dependence, and such a model may also be extended to multi-level models by the current perturbation approach.

\section*{Acknowledgements}
%We are grateful for helpful discussions with ...
This work was supported by the National Natural Science Foundation of China under Grant No. 12105094, and by the Fundamental Research Funds for the Central Universities from China.

%\newpage
\clearpage

\widetext
\begin{center}

\textbf{\large Supplemental Material for ``Perturbative approach to time-dependent quantum systems and applications to one-crossing multistate Landau-Zener models''}
\end{center}

\section*{Appendix A: Derivation of the probabilities for one-crossing MLZ models}

\setcounter{figure}{0}
\setcounter{equation}{0}
\setcounter{subsubsection}{0}
\renewcommand{\theequation}{A\arabic{equation}}
\renewcommand\thefigure{A\arabic{figure}}

In this Appendix, we are going to calculate explicitly the matrix elements of %$W_n(\infty)$ and
$P_n$ of the general one-crossing MLZ model (the model (1) with $E(t)$ given by (35) in the main text) up to $n=4$, namely, transition probabilities up to $4$th order in the couplings. Our calculation procedures are presented in a sufficiently detailed manner so readers interested in reproducing them can readily do so.

Note that due to the symmetry properties (30) and (31) in the main text, it suffices to determine those elements $P_{n,jk}$ with $j\le k$.  %Readers not interested in these details can skip directly to Subsection C for the final results of $P_n$ up to $n=4$.

\subsubsection{Probabilities up to $g^3$ }

We start by writing out the matrix elements of $W_1(t)$, $W_2(t)$, and $W_3(t)$ explicitly:
\begin{align}\label{}
&W_{1,jk}(t)=-2i\int_0^{t}  ds  \tilde A_{jk} (s),\\%= -2i \int_0^t ds e^{i (\phi_j(s)-\phi_k(s))}  A_{jk} ,\\
&W_{2,jk}(t)=-2\left\{\left[ \int_0^{t}  ds  \tilde A (s)\right]^2\right\}_{jk} ,\\
%&=-2 \sum_l [\int_0^t ds e^{i (\phi_j(s)-\phi_l(s))}  A_{jl}][  \int_0^t ds e^{i (\phi_l(s)-\phi_k(s))}  A_{lk}] ,\\
%&=-2 \sum_l [\int_0^t ds    \tilde A_{jl}(s)][\int_0^t ds    \tilde A_{lk}(s)] ,\\
&W_{3,jk}(t)= 2i\left\{\left[\int_0^t ds  \tilde A  (s) \right]^3 \right\}_{jk}
 % &+2i\{\int_0^t ds  [\int_0^{s } ds_1 \tilde A  (s_1)] \tilde A (s) [\int_0^{s } ds_1 \tilde A  (s_1)]\}_{jk}\nn\\
%&=-2 i \sum_{l,p} \{[\int_0^t ds    \tilde A_{jl}(s)][\int_0^t ds    \tilde A_{lp}(s)] [\int_0^t ds    \tilde A_{pk}(s)] \}  \nn\\
 -2i \sum_{l,p} \int_0^t ds  \left[\int_0^{s } ds_1 \tilde A_{jl}  (s_1)\right] \tilde A_{lp}   (s) \left[\int_0^{s } ds_1 \tilde A_{pk}    (s_1)\right] .
\end{align}
Recall that $\tilde A_{jj} (t) = 0$ and $ \tilde A_{jk} (t) = e^{i (\phi_j(t)-\phi_k(t))}  A_{jk}$ for $j\ne k$. The phase $\phi_j(t)$ reads:
\begin{align}\label{}
\phi_j(t)=\int_{0}^t ds b_j s=\frac{1}{2}b_j t^2.
\end{align}
So %for $b_j\ne b_k$ we have
\begin{align}\label{eq:intA}
&\int_0^t ds \tilde A_{jk} (s) =  A_{jk}  \int_0^t ds  e^{i \frac{1}{2} b_{jk}s^2}
%&= A_{jk}  \int_0^t ds[ \cos( \frac{1}{2} b_{jk} s^2)+i  \sin( \frac{1}{2} b_{jk} s^2)]\nn\\
=  A_{jk} \sqrt{\frac{\pi}{|b_{jk}|}}\left[ C\left( \sqrt{\frac{|b_{jk}|}{\pi}} t\right)+i\sgn(b_{jk}) S\left( \sqrt{\frac{|b_{jk}|}{\pi}} t\right)\right],
\end{align}
where $\sgn$ is the sign function, the slope differences are denoted  as
\begin{align}\label{}
b_{jk}\equiv b_j-b_k
\end{align}
for notation simplicity, and $C(x)$ and $S(x)$ ($x\ge0$) are the Fresnel integrals \cite{Abramowitz-Stegun} defined as
\begin{align}\label{}
&C(x)=\int_0^x\cos \left(\frac{\pi}{2}t^2\right)dt,\\
&S(x)=\int_0^x\sin \left(\frac{\pi}{2}t^2\right)dt.
\end{align}

Let's now determine the matrix elements of $P_n$ up to $n=3$. For this purpose, it suffices to consider $W_1(\infty)$ and  $W_2(\infty)$. Using limits of the Fresnel integrals at infinity
\begin{align}\label{}
C(\infty)=S(\infty)= \frac{1}{2},
\end{align}
one arrives at
\begin{align}\label{eq:Atilde-integral}
&\int_0^\infty ds \tilde A_{jk} (s) %=    \frac{1}{2}\sqrt{\frac{\pi}{|b_{jk}|}}\left[ 1+i\sgn(b_{jk})  \right]A_{jk}\nn\\
= \sqrt{\frac{\pi}{2|b_{jk}|}}e^{i\frac{\pi}{4}\sgn(b_{jk})} A_{jk}
=  \frac{1}{\sqrt 2 } \lambda_{jk}e^{i\frac{\pi}{4}\sgn(b_{jk})},
\end{align}
where we defined the combinations
\begin{align}\label{eq:lambdajk}
\lambda_{jk}=\sqrt{\frac{\pi}{|b_{jk}|}}A_{jk}, \quad \textrm{for } j\ne k.
\end{align}
These combinations will appear frequently during our analysis. Later, whenever possible, expressions  will be written in terms of these $\lambda_{jk}$. From \eqref{eq:Atilde-integral}, $W_{1}(\infty)$ follows immediately:
\begin{align}\label{}
&W_{1,jj}(\infty)=0,\\
&W_{1,jk}(\infty)=-i\sqrt{2}  \lambda_{jk} e^{i\frac{\pi}{4}},  \quad \textrm{for } j< k.
\end{align}
$W_{2}(\infty)$ involves a sum
\begin{align}\label{}
&W_{2,jk}(\infty)%=-2 \sum_l  \sqrt{\frac{\pi}{2|b_{jl}|}} e^{i\frac{\pi}{4}\sgn(b_{jl})}  A_{jl} \nn\\
%&\times \sqrt{\frac{\pi}{2|b_{lk}|}} e^{i\frac{\pi}{4}\sgn(b_{lk})}  A_{lk}  \nn\\
%&=-\frac{\pi}{2} \sum_l \left\{ \sqrt{\frac{1}{|b_{jl}|}}\left[ 1+i\sgn(b_{jl})  \right]A_{jl}\right\}\nn\\
%&\times\left\{ \sqrt{\frac{1}{|b_{lk}|}}\left[ 1+i\sgn(b_{lk})  \right]A_{lk}\right\} \nn\\
=- \sum_l  \lambda_{jl}  \lambda_{lk} e^{i\frac{\pi}{4}[\sgn(b_{jl})+\sgn(b_{lk})]}.
\end{align}
Let's look at the diagonal and off-diagonal elements of  $W_{2}(\infty)$ separately. For $j=k$ one always has $\sgn(b_{jl})=-\sgn(b_{lk})$, so
\begin{align}\label{}
&W_{2,jj}(\infty)%=-\frac{\pi}{2} \sum_l  \frac{A_{jl} A_{lj}}{\sqrt{{|b_{jl}b_{lj}|} }} [1-(-1)] \nn\\
=-  \sum_{l}^{l\ne j}   \lambda_{jl}^2   .
\end{align}
For $j< k$, if  $l<j$ or  $l>k$ we have $\sgn(b_{jl})=-\sgn(b_{lk})$, and if $j<l<k$ we have $\sgn(b_{jl})=\sgn(b_{lk})=1$, %and if $k<l<j$ we have $\sgn(b_{jl})=\sgn(b_{lk})=-1$,
so
\begin{align}\label{}
&W_{2,jk}(\infty)%=-\pi\left[\sum_{l}^{ l<j\textrm{ or }  l>k    } \frac{A_{jl} A_{lk}}{\sqrt{{|b_{jl}b_{lk}|} }} +\sum_{l}^{j<l<k  } \frac{A_{jl} A_{lk}}{\sqrt{{|b_{jl}b_{lk}|} }}  i \right]\nn\\
=-  \sum_{l}^{ l<j\textrm{ or }  l>k   }  \lambda_{jl}  \lambda_{lk}  -i  \sum_{l}^{j<l<k } \lambda_{jl}  \lambda_{lk} . %,  \quad \textrm{for } j< k.
\end{align}
\begin{comment}
Summarizing, we obtain $W(\infty)$ up to second order in $g$ as:
\begin{align}\label{}
&W_{jj}(\infty)=1-\pi g^2 \sum_{l}^{l\ne j}  \frac{A_{jl}^2}{|b_{jl}|} +O(g^3),
\end{align}
and for $j \ne k$:
\begin{align}\label{}
&W_{jk}(\infty)=\left[ \sgn(b_{jk})-i  \right]\sqrt{\frac{\pi}{|b_{jk}|}} g A_{jk} \nn\\
& -\pi g^2\sum_{l}^{l>\max(j,k)\textrm{ or }  l<\min(j,k) } \frac{A_{jl} A_{lk}}{\sqrt{{|b_{jl}b_{lk}|} }}   \nn\\
&-i\pi g^2\sgn(b_{jk}) \sum_{l}^{\min(j,k)<l<\max(j,k)  } \frac{A_{jl} A_{lk}}{\sqrt{{|b_{jl}b_{lk}|} }}  +O(g^3).
\end{align}
\end{comment}
From $W_1(\infty)$ and $W_2(\infty)$, the transition probabilities can then be determined up to terms with $g^3$. The diagonal elements are:
\begin{align}\label{eq:Pdiag-upto3}
&P_{jj} =|W_{jj}(\infty)|^2=1-2  g^2 \sum_{l}^{l\ne j} \lambda_{jl}^2 +O(g^4).
\end{align}
The off-diagonal elements with $j< k$ are:
\begin{align}\label{eq:Poffdiag-upto3}
&P_{jk}=|W_{jk}(\infty)|^2=%|\left[ \sgn(b_{jk})-i  \right]\sqrt{\frac{\pi}{|b_{jk}|}} g A_{jk} \nn\\
%& -\pi g^2\sum_{l}^{l>\max(j,k)\textrm{ or }  l<\min(j,k) } \frac{A_{jl} A_{lk}}{\sqrt{{|b_{jl}b_{lk}|} }}   \nn\\
%&-i\pi g^2\sgn(b_{jk}) \sum_{l}^{\min(j,k)<l<\max(j,k)  } \frac{A_{jl} A_{lk}}{\sqrt{{|b_{jl}b_{lk}|} }} |^2 +O(g^4)\nn\\
2 g^2  \lambda_{jk}^2   -2 g^3  \lambda_{jk} \left(\sum_{l}^{ l<j\textrm{ or }  l>k    } \lambda_{jl}  \lambda_{lk}-\sum_{l}^{j<l<k  } \lambda_{jl}  \lambda_{lk} \right) +O(g^4).
\end{align}

\subsubsection{Probabilities in $g^4$ }
Below we determine terms in $g^4$ for  the transition probabilities, which analysis is more complicated. Let's first consider the off-diagonal elements $P_{4,jk}$ with $j<k$. We need to calculate $W_{3,jk}(\infty)$, which reads
\begin{align}\label{eq:W3}
&W_{3,jk}(\infty)= 2i\left\{\left[\int_0^\infty ds  \tilde A  (s) \right]^3 \right \}_{jk}
-2i \sum_{l,p} \int_0^\infty ds  \left[\int_0^{s } ds_1 \tilde A_{jl}  (s_1)\right] \tilde A_{lp}   (s) \left[\int_0^{s } ds_1 \tilde A_{pk}    (s_1)\right].
\end{align}
The integrals in the first term in \eqref{eq:W3}  can be evaluated similarly as $W_{1,jk}(\infty)$ and $W_{2,jk}(\infty)$ using the limits of Fresnel integrals:
\begin{align}\label{eq:W3-1st}
&\left\{\left[\int_0^\infty ds  \tilde A  (s) \right ]^3 \right\}_{jk}
%&=\sum_{l,p}  [\int_0^\infty ds    \tilde A_{jl}(s)][\int_0^\infty ds    \tilde A_{lp}(s)] [\int_0^\infty ds    \tilde A_{pk}(s)]  \nn\\
= \frac{1}{2\sqrt 2}\sum_{l,p}  \lambda_{jl}\lambda_{lp}\lambda_{pk} e^{i\frac{\pi}{4}[\sgn(b_{jl})+\sgn(b_{lp})+\sgn(b_{pk})]}
=\left(\frac{\pi}{2}\right)^\frac{3}{2}\sum_{l,p}   \frac{ A_{jl}A_{lp}A_{pk}}{ \sqrt {-i b_{jl}}\sqrt {-i b_{lp}}\sqrt {-i b_{pk}}  },
\end{align}
where in the last equality we used the fact that $b_{jl}$ is a non-zero real number, so
\begin{align}\label{eq:sqrt-b}
\sqrt{-i b_{jl} }=\sqrt{|b_{jl}|}e^{-i\frac{\pi}{4} \sgn(b_{jl}) },
\end{align}
and similarly for $b_{lp}$ and $b_{pk}$ (the branch cut of a square root function is taking to be slightly below the negative real axis).

Evaluation of the second  term in \eqref{eq:W3}  takes much more efforts. It is a sum of terms of the form
\begin{align}\label{}
&\int_0^\infty ds \tilde A_{lp}   (s)  \int_0^{s } ds_1 \tilde A_{jl}  (s_1) \int_0^{s } ds_2 \tilde A_{pk}    (s_2)
 \equiv A_{jl}   A_{lp}   A_{pk}  Q,
\end{align}
where $Q$ is the following integral:
\begin{align}\label{eq:Q-def}
&Q=\int_0^\infty ds  e^{i\frac{1}{2}b_{lp}  s^2}  \int_0^{s } ds_1 e^{i\frac{1}{2}b_{jl}  s_1^2} \int_0^{s } ds_2 e^{i\frac{1}{2}b_{pk}  s_2^2} \nn\\
&=\frac{ \pi  }{\sqrt{|b_{jl}b_{pk}|}} \int_0^\infty ds  e^{i\frac{1}{2}b_{lp}  s^2}
 \left[ C\left( \sqrt{\frac{|b_{jl}|}{\pi}} s\right)+i\sgn(b_{jl}) S\left( \sqrt{\frac{|b_{jl}|}{\pi}} s\right)\right]
\left[ C\left( \sqrt{\frac{|b_{pk}|}{\pi}} s\right)+i\sgn(b_{pk}) S\left( \sqrt{\frac{|b_{pk}|}{\pi}} s\right)\right].
\end{align}
We find that result of this integral separate into two cases depending on whether the values of the slopes satisfy at least one of $b_{jl}+b_{lp}=0$ and  $b_{lp}+b_{pk}=0$. We call this condition the ``resonance condition''. The case that this condition is not satisfied will be named the ``off-resonant'' case, and the other case that this condition is satisfied will be named the ``resonant'' case. Note that $b_{jl}+b_{lp}=0$ means $p=j$ and  $b_{lp}+b_{pk}=0$ means $l=k$, so the resonance condition is equivalent to saying that at least one of $p=j$ and $l=k$ is satisfied. %at least one same index appear more than once in the path  $j\rar l \rar p\rar k$...

In the off-resonant case, this integral \eqref{eq:Q-def} can be evaluated using the parametric integration technique. The procedure is rather complicated so details will be given in a separate Appendix B, and here we present directly the result:
\begin{align}\label{Q-result-off-res}
&Q=\sqrt{\frac{\pi}{2}}\frac{1}{  \sqrt{-i b_{jl} } \sqrt{-i b_{lp} } \sqrt{-i b_{pk}  } }  %\nn\\
 \left[ \theta (- b_{lp})\theta( b_{jp})\theta (b_{lk})   \pi+ \Arctan  \frac{ \sqrt{-i b_{jl} }  \sqrt{-i b_{pk} }  }{ \sqrt{-i b_{lp} } \sqrt{-ib_{jk}}} \right] ,
\end{align}
where $\theta(x)$ is the Heaviside step function, and Arctan is the complex-valued inverse tangent function defined in the principal branch (the branch cuts are taken to lie on the imaginary axis at $(-i\infty,-i]$ and $[-i,i\infty)$; values of Arctan have real parts $ \pi/2$ or $ -\pi/2$ for points on the upper/lower branch cut, respectively).

In the resonant case,  this integral \eqref{eq:Q-def} diverges. This means that the corresponding term in $ W_{3,jk}(t)$ does not have a limit at $t\rar \infty$. However, we recall that the probability $P_{jk}$, what we are interested in, is bounded between $0$ and $1$ at any value of $g$, and we expect that it should have a well-defined limit at each order of $g$. Thus, $P_{jk}(\infty)$ must converge even if $ W_{jk}(\infty)$ does not. Let's look at $P_{4,jk}$ with $j<k$. Since $W_{0,jk}(t)$ vanishes, according to (21) in the main text, the leading order in which $W_{3,jk}$ appears is $P_{4,jk}$, which reads
\begin{align}\label{}
&P_{4,jk}=\lim_{t\rar\infty}  \{|W_{2,jk}(t)|^2 +[W_{1,jk}^*(t)  W_{3,jk}(t)+c.c.]\}.
\end{align}
Thus, a resonant term contributes to $P_{4,jk}$ via the combination
\begin{align}\label{}
&R\equiv\lim_{t\rar\infty}\left\{\left[ -2i\int_0^{t}  ds  \tilde A_{jk} (s)  \right]^* \right.
\left. \left[-2i\int_0^t ds \tilde A_{lp}   (s)  \int_0^{s } ds_1 \tilde A_{jl}  (s_1) \int_0^{s } ds_2 \tilde A_{pk}    (s_2)\right]+c.c.\right\}.
%&= 4\pi^{\frac{3}{2}} \frac{ A_{jk}  A_{jl}  A_{lp}  A_{pk}  }{\sqrt{|b_{jk}b_{jl} b_{pk}|}} \lim_{t\rar\infty} \left[ C\left( \sqrt{\frac{|b_{jk}|}{\pi}} t\right)- i S\left( \sqrt{\frac{|b_{jk}|}{\pi}} t\right)\right]\nn\\
%&\times    \int_0^t ds  e^{i \frac{1}{2} b_{lp}s^2} \left[  C\left( \sqrt{\frac{|b_{jl}|}{\pi}} s\right)+i\sgn(b_{jl}) S\left( \sqrt{\frac{|b_{jl}|}{\pi}} s\right)\right]\nn\\
%&\times\left[  C\left( \sqrt{\frac{|b_{pk}|}{\pi}} s\right)+i\sgn(b_{pk}) S\left( \sqrt{\frac{|b_{pk}|}{\pi}} s\right)\right]+c.c.
\end{align}
Calculating this limit  numerically %(in terms of Fresnel integrals) at large $t$
for different choices $b_{jl}$, $b_{lp}$ and $b_{pk}$ satisfying the resonance condition, we find that it indeed converges, and the limit is
\begin{comment}
\begin{align}\label{eq:resonant}
&\left[-i\sqrt{\frac{2\pi}{|b_{jk}|}} e^{i\frac{\pi}{4}} A_{jk}\right]^* (-2i)A_{jl}   A_{lp}   A_{pk}  Q+c.c.\nn\\
&=2 \sqrt{\frac{ \pi}{|b_{jk}|}}  A_{jk}  A_{jl}   A_{lp}   A_{pk}( 2\operatorname{Re} Q+ 2\operatorname{Im} Q)\nn\\
&=4\sqrt{\frac{ \pi}{|b_{jk}|}}  A_{jk} A_{jl}   A_{lp}   A_{pk} \frac{ \pi  }{\sqrt{|b_{jl}b_{pk}|}} \sqrt{\frac{ 2}{|b_{lp}|}} \frac{\sqrt\frac{\pi}{2}}{4} \nn\\
&=  \pi^2 \frac{ A_{jk}  A_{jl}   A_{lp}   A_{pk}  }{\sqrt{|b_{jk}b_{jl}b_{lp}b_{pk}|}}.
\end{align}
\end{comment}
\begin{align}\label{eq:resonant}
R=\left\{\begin{array}{cc}
              %    \pi^2 \frac{  A_{jk}^2 A_{jl}^2   }{|b_{jk}|\sqrt{b_{jl}b_{lp}b_{pk}|}}, &   \textrm{for either $p= j$ or $l=k$,} \\
                \sgn(b_{lk}) \lambda_{jk}^2 \lambda_{jl}^2    , &   \textrm{for $p= j$ and $l\ne k$,} \\
               \sgn(b_{jp}) \lambda_{jk}^2 \lambda_{pk}^2  , &   \textrm{for $p\ne j$ and $l= k$,} \\
                  0, & \textrm{for $p=j$ and $l=k$.}
                \end{array}\right.
\end{align}

All other contributions to $P_{4,jk}$ can be calculated directly using the results obtained above for $W_{1,jk}(\infty)$, $W_{2,jk}(\infty)$, and other terms in $W_{3,jk}(\infty)$. The term  $|W_{2,jk}(\infty)|^2$ reads:
\begin{align}\label{}
&|W_{2,jk}(\infty)|^2=  \left( \sum_{l}^{ l<j\textrm{ or }  l>k   }  \lambda_{jl} \lambda_{lk}\right)^2 +\left( \sum_{l}^{j<l<k }  \lambda_{jl} \lambda_{lk} \right)^2 .
\end{align}
The first term in \eqref{eq:W3}, namely the completely symmetric term, gives a contribution:
\begin{align}\label{eq:symmetric}
&P_{4,jk}^{sym}
%&=\left(-i\sqrt{\frac{2\pi}{|b_{jk}|}} e^{i\frac{\pi}{4}} A_{jk}\right)^*(  2i)\left(\frac{\pi}{2}\right)^\frac{3}{2}\sum_{l,p}   \frac{ A_{jl}A_{lp}A_{pk}}{ \sqrt {-i b_{jl}}\sqrt {-i b_{lp}}\sqrt {-i b_{pk}}  }+c.c.\nn\\
=-  \lambda_{jk}  \sum_{l,p}   \lambda_{jl} \lambda_{lp} \lambda_{pk} e^{i\frac{\pi}{4}[\sgn(b_{jl})+\sgn(b_{lp})+\sgn(b_{pk})-1]}+c.c.
\end{align}
The quantity $\sgn(b_{jl})+\sgn(b_{lp})+\sgn(b_{pk})-1$ can take $3$ possible values: $0$, and $\pm 2$. When added with the complex conjugated terms, only terms with $\sgn(b_{jl})+\sgn(b_{lp})+\sgn(b_{pk})-1=0$ survive, so
\begin{align}\label{eq:symmetric-2}
&P_{4,jk}^{sym}=-2  \lambda_{jk} \sum_{l,p}^{\sgn(b_{jl})+\sgn(b_{lp})+\sgn(b_{pk})=1}    \lambda_{jl} \lambda_{lp} \lambda_{pk}.
\end{align}
The off-resonance terms give a contribution:
\begin{align}\label{eq:P4-off-res}
&P_{4,jk}^{off-res}%\nn\\
%&=\left(-i\sqrt{\frac{2\pi}{|b_{jk}|}} e^{i\frac{\pi}{4}} A_{jk}\right)^*( - 2i)\sqrt\frac{\pi}{2} \sum_{l,p}^{p\ne j,l\ne k }   \frac{ A_{jl}A_{lp}A_{pk}}{ \sqrt {-i b_{jl}}\sqrt {-i b_{lp}}\sqrt {-i b_{pk}}  } \nn\\
%&\times\left[ \theta (- b_{lp})\theta( b_{jp})\theta (b_{lk})   \pi+ \Arctan  \frac{ \sqrt{-i b_{jl} }  \sqrt{-i b_{pk} }  }{ \sqrt{-i b_{lp} } \sqrt{-ib_{jk}}} \right]+c.c. \nn\\
=2\lambda_{jk}
\sum_{l,p}^{p\ne j,l\ne k }   \lambda_{jl} \lambda_{lp} \lambda_{pk} e^{i\frac{\pi}{4}[\sgn(b_{jl})+\sgn(b_{lp})+\sgn(b_{pk})-1]}
\left[ \theta (- b_{lp})\theta( b_{jp})\theta (b_{lk})   \pi+ \Arctan  \frac{ \sqrt{-i b_{jl} }  \sqrt{-i b_{pk} }  }{ \sqrt{-i b_{lp} } \sqrt{-ib_{jk}}} \right]+c.c.
\end{align}
The phase of the argument of Arctan function is
\begin{align}\label{eq:arg-Arctan}
&\arg \frac{ \sqrt{-i b_{jl} }  \sqrt{-i b_{pk} }  }{ \sqrt{-i b_{lp} } \sqrt{-ib_{jk}}}=\frac{\pi}{2}\sgn(b_{lp})%=\frac{\pi}{4}[-\sgn(b_{jl})+\sgn(b_{lp})-\sgn(b_{pk})+1]
 -\frac{\pi}{4}[\sgn(b_{jl})+\sgn(b_{lp})+\sgn(b_{pk})-1].
\end{align}
When $\sgn(b_{jl})+\sgn(b_{lp})+\sgn(b_{pk})-1=\pm2$, the exponential phase factor in \eqref{eq:P4-off-res}  is $\pm i$, so the imaginary part of the Arctan function is selected. But according to \eqref{eq:arg-Arctan}, $\arg[ \sqrt{-i b_{jl} }  \sqrt{-i b_{pk} }  /(\sqrt{-i b_{lp} } \sqrt{-ib_{jk}})]$ can only be $0$ or $\pm \pi$, and the Arctan function is always real, so the contribution vanishes. When $\sgn(b_{jl})+\sgn(b_{lp})+\sgn(b_{pk})-1=0$, the real part of the Arctan function is selected. In this case,
\begin{align}\label{}
&\arg \frac{ \sqrt{-i b_{jl} }  \sqrt{-i b_{pk} }  }{ \sqrt{-i b_{lp} } \sqrt{-ib_{jk}}}=\frac{\pi}{2}\sgn(b_{lp}),
\end{align}
and for a real $x$,
\begin{align}\label{}
 \operatorname{Re}[\Arctan(i x)]=\left\{\begin{array}{cc}
                \frac{\pi}{2} \sgn(x), &   \textrm{for $|x|>1$,} \\
                  0, & \textrm{for $|x|<1$.}
                \end{array}\right.
\end{align}
We thus obtain
\begin{align}\label{}
 \operatorname{Re}\left[\Arctan  \frac{ \sqrt{-i b_{jl} }  \sqrt{-i b_{pk} }  }{ \sqrt{-i b_{lp} } \sqrt{-ib_{jk}}}\right] =\frac{\pi}{2} \theta( -b_{jp} b_{lk})\sgn(b_{lp}).
\end{align}
Therefore, the off-resonance contribution reads:
\begin{align}\label{}
&P_{4,jk}^{off-res}=4 \lambda_{jk} \sum_{l,p,p\ne j,l\ne k }^{\sgn(b_{jl})+\sgn(b_{lp})+\sgn(b_{pk})=1}    \lambda_{jl}\lambda_{lp}\lambda_{pk}
\left[ \theta (- b_{lp})\theta( b_{jp})\theta (b_{lk})   +\frac{1}{2} \theta( -b_{jp} b_{lk})\sgn(b_{lp})\right].
\end{align}
One sees that though the Arctan function appears in the expression of the integral \eqref{Q-result-off-res}, it does not enter the expression of $P_{4,jk}$.
\begin{comment}
\begin{align}\label{}
& \theta (- b_{lp})\theta( b_{jp})\theta (b_{lk})   +\frac{1}{2} \theta( -b_{jp} b_{lk})\sgn(b_{lp}) -\frac{1}{2}\nn\\
&= \theta (- b_{lp})\theta( b_{jp})\theta (b_{lk})   +\frac{1}{2} \theta( -b_{jp} b_{lk})[-2\theta(-b_{lp} )+1]-\frac{1}{2}\nn\\
&= \theta (- b_{lp})[\theta( b_{jp})\theta (b_{lk}) -\theta( -b_{jp} b_{lk})] - \frac{1}{2}\theta( b_{jp} b_{lk}) \nn\\
&\rar (j<p<l<k )-(p<j<l<k)- (j<p<k<l  )  \nn\\
&- \frac{1}{2}[(j<p<l<k )+jlkp+ljpk+(p<j<k<l )]\nn\\
&=-[ pjlk+ jpkl + \frac{1}{2}(-jplk +jlkp+ljpk+pjkl).]
\end{align}
\end{comment}

Finally, we collect all contributions to $P_{4,jk}$. In $P_{4,jk}^{sym}$ we separate the terms with $p= j$ or $l=k$, which can be combined with the resonance contribution; the rest terms can be combined with the off-resonance contribution. After simplifications, this gives the final expression of the $4$th order contribution to the probability $P_{jk}$ for $j<k$:
\begin{align}\label{eq:P4jk}
&P_{4,jk}= \left( \sum_{l}^{ l<j\textrm{ or }  l>k   }  \lambda_{jl} \lambda_{lk} \right)^2  + \left( \sum_{l}^{j<l<k }  \lambda_{jl} \lambda_{lk}  \right)^2
- \lambda_{jk}^2\left(\sum_{l}^{l<k}     \lambda_{jl}^2 +3\sum_{l}^{l>k}  \lambda_{jl}^2  +\sum_{l}^{l>j}   \lambda_{lk}^2  +3\sum_{l}^{l<j} \lambda_{lk}^2  +2  \lambda_{jk}^2   \right) \nn\\
%&\left.-4 \frac{A_{jk}}{\sqrt{|b_{jk}|}}\sum_{l,p  }^{p<j<l<k\textrm{ or }  j<p<k<l \textrm{ or }  p<j<k<l}   \frac{ A_{jl}A_{lp}A_{pk}}{ \sqrt {| b_{jl}  b_{lp}  b_{pk}|}  } \right].
&-2 \lambda_{jk} \left(2\sum_{l,p  }^{p<j<l<k\textrm{ or }  j<p<k<l } -\sum_{l,p  }^{ j<p<l<k } \right.
 \left.+ \sum_{l,p  }^{j<l<k<p\textrm{, }  l<j<p<k \textrm{ or }  p<j<k<l}\right)\lambda_{jl}\lambda_{lp}\lambda_{pk}  .
\end{align}
%where we used %in the final sum of $l$ and $p$
%\begin{align}\label{}
%&\sum_{l,p,p\ne j,l\ne k }^{\sgn(b_{jl})+\sgn(b_{lp})+\sgn(b_{pk})=1}   \theta (- b_{lp})[\theta( b_{jp})\theta (b_{lk})  -1]\nn\\
%&=-\sum_{l,p  }^{p<j<l<k\textrm{ or }  j<p<k<l \textrm{ or }  p<j<k<l}  1.
%\end{align}

Since the matrix $P_4$ is symmetric, %according to \eqref{eq:Pn-even},
the off-diagonal elements $P_{4,jk}$ with $j>k$ are connected to those in \eqref{eq:P4jk} by
\begin{align}\label{}
P_{4,jk}=P_{4,kj},\quad \textrm{for } j>k.
\end{align}
The diagonal elements can then be obtained from the doubly stochastic property of the  probability matrix $P$.
%\begin{align}\label{}
%P_{4,jj}=1-\sum_{k}^{k\ne j}P_{4,jk}.
%\end{align}
We then arrive at the result presented in Section IIIA of the main text.

Note that the probabilities up to $4$th order in the couplings can be expressed as polynomials of $\lambda_{jk}$ with integer coefficients. One may wonder if the same is true for higher order terms of any one-crossing MLZ model, but it turns out that this is not the case. For example, in the one-crossing $5$-state LZ model discussed in \cite{Hu-5-state} (the model in Section IVB3 of the main text), one finds from numerics that the higher order terms depend not only on $A_{jk}/\sqrt{|b_{jk}|}$ but also on the ratios of the slope differences $b_{jk}/b_{pl}$. In fact, in the integral \eqref{Q-result-off-res} for the off-resonant case, the Arctan function does depend on the ratios of the slope differences. Although this Arctan function finally does not enter the $4$th order terms, it is possible that it does enter higher order terms; there may also be other forms of functions arising from higher-dimensional multiple integrations which also depend on $b_{jk}/b_{pl}$.

\section*{Appendix B: integral $Q$ in the off-resonant case}

\setcounter{figure}{0}
\setcounter{equation}{0}
\renewcommand{\theequation}{B\arabic{equation}}
\renewcommand\thefigure{B\arabic{figure}}

In this appendix, we present details of calculation of the integral $Q$ in Eq.~\eqref{eq:Q-def} in the off-resonant case.

For notation simplicity let's write $\alpha=b_{lp}/2$, $\beta=b_{jl}/2$ and $\gamma=b_{pk}/2$, so \eqref{eq:Q-def} becomes
\begin{align}\label{eq:Q}
Q =\int_0^\infty ds  e^{i\alpha s^2}   \int_0^{s } ds_1e^{i\beta s_1^2}   \int_0^{s } ds_2e^{i\gamma s_2^2}   .
\end{align}
Note the order of the three quantities defined. The resonance condition then reads $\alpha+\beta=0$ or $\alpha+\gamma=0$. Note also that we have $\alpha+\beta+\gamma> 0$ since we are considering off-diagonal elements of $W_n(\infty)$ with $j<k$.

As written in \eqref{eq:Q-def}, %The two inner integrals in \eqref{eq:Q} can still be expressed as combinations of Fresnel integrals;
$Q$ can be expressed as an  integral of $ e^{i\alpha s^2} $ multiplied by  two Fresnel integrals with arguments being generally different. An integral of this form  cannot be evaluated directly (e.g. using Mathematica). Nevertheless, it can be performed using the parametric integration technique. We first make changes of variables $s= \sqrt u$, $s_1= \sqrt u_1$ and $s_2= \sqrt u_2$, so
\begin{align}\label{}
Q =\int_0^\infty du \frac{ e^{i\alpha u} }{2\sqrt u}  \int_0^{u } du_1 \frac{ e^{i\beta u_1} }{2\sqrt u_1} \int_0^{u } du_2\frac{ e^{i\gamma u_2} }{2\sqrt u_2}  .
\end{align}
We now replace the square roots by using Gaussian integrals of new parameters $x$, $x_1$ and $x_2$ as:
\begin{align}\label{}
\frac{1}{2\sqrt u}= \frac{1}{\sqrt {\pi}}\int_0^\infty dx e^{-u x^2},
\end{align}
and similarly for $\sqrt u_1$ and $\sqrt u_2$, then
\begin{align}\label{}
&Q %=\frac{1}{\pi^{\frac{3}{2}}}\int_0^\infty du  e^{i\alpha u}   \int_0^\infty dx e^{-u x^2} (\int_0^{u } du_1   e^{i\beta u_1}   \int_0^\infty dx_1 e^{-u_1 x_1^2}  ) \nn\\
%& \times(\int_0^{\infty } du_2e^{i\gamma u_2}  \int_0^\infty dx_2 e^{-u_2 x_2^2}   )\nn\\
=\frac{1}{\pi^{\frac{3}{2}}}\int_0^\infty du  \int_0^\infty dx e^{(-x^2+i\alpha) u}     \int_0^{u } du_1 \int_0^\infty dx_1   e^{(-x_1^2 +i\beta )u_1 }    \int_0^{u } du_2 \int_0^\infty dx_2   e^{(-x_2^2 +i\gamma )u_2 }    .
\end{align}
Switching the order of integrations and performing the integrals on $u_1$, $u_2$, and then $u$, we get
\begin{align}\label{}
&Q %=\frac{1}{\pi^{\frac{3}{2}}} \int_0^\infty dx\int_0^\infty du  e^{(-x^2+i\alpha) u}     \int_0^\infty dx_1  \frac{ e^{(-x_1^2 +i\beta )u }-1}{-x_1^2 +i\beta }   \nn\\
%& \times  \int_0^\infty dx_2 \frac{ e^{(-x_2^2 +i\gamma )u }-1}{-x_2^2 +i\gamma }     \nn\\
=\frac{1}{\pi^{\frac{3}{2}}} \int_0^\infty dx\int_0^\infty dx_1 \int_0^\infty dx_2  \frac{1}{\left(x_1^2-i \beta \right) \left(x_2^2-i \gamma \right)}\nn\\
&\times\left[  \frac{1}{x^2-i \alpha }-\frac{1}{x^2+x_1^2-i (\alpha +\beta )}-\frac{1}{x^2+x_2^2-i (\alpha +\gamma )}\right.
\left.+\frac{1}{x^2+x_1^2+x_2^2-i (\alpha +\beta +\gamma )}\right]\equiv Q_1+Q_2+Q_3+Q_4.
\end{align}
%For a general $\alpha$, $\beta$ and $\gamma$, integrals of each of these $4$ terms can be performed; the details are presented in Appendix A, and here...  we present the result:

Below we evaluate the $4$ terms separately. The first term $Q_1$ separates to three single integrals which can be readily performed:
\begin{align}\label{}
&Q_1  %=\frac{1}{\pi^{\frac{3}{2}}}\frac{\pi }{2 \sqrt{-i \alpha }} \frac{\pi }{2 \sqrt{-i \beta }} \frac{\pi }{2 \sqrt{-i \gamma }}.
=\frac{\pi^{\frac{3}{2}}e^{i\frac{\pi}{4}[ \sgn(\alpha)+\sgn(\beta)+\sgn(\gamma)] }}{8 \sqrt{|\alpha \beta \gamma|}}
=\frac{\pi^{\frac{3}{2}} }{8 \sqrt{-i \alpha }\sqrt{-i \beta }\sqrt{-i \gamma } },
\end{align}
where in obtaining the second equality recall that the branch cut of a square root function is taking to be slightly below the negative real axis, and that $\alpha$, $\beta$ and $\gamma$ are all real and non-zero, so we have
\begin{align}
&\sqrt{-i \alpha }=\sqrt{|\alpha|}e^{-i\frac{\pi}{4} \sgn(\alpha) },\label{eq:sqrt}\\
& \sqrt{ \alpha }=\sqrt{|\alpha|} e^{i\frac{\pi}{4}[1- \sgn(\alpha)] },
\end{align}
and similarly for $\beta$ and $\gamma$. %So $Q_1$ can also be written as:

The integrals in $Q_2$ can also be performed directly by first integrating $x_2$ and $x$ and then $x_1$:
\begin{align}
&Q_2 %=-\frac{1}{\pi^{\frac{3}{2}}} \int_0^\infty dx\int_0^\infty dx_1 \int_0^\infty dx_2  \frac{1}{\left(x_1^2-i \beta \right) \left(x_2^2-i \gamma \right)}\nn\\
%&\times   \frac{1}{x^2+x_1^2-i (\alpha +\beta )}  \nn\\
%&=-\frac{1}{\pi^{\frac{3}{2}}} \frac{\pi }{2 \sqrt{-i \gamma }}  \int_0^\infty dx\int_0^\infty dx_1  \frac{1}{\left(x_1^2-i \beta \right)}\nn\\
%&\times   \frac{1}{x^2+x_1^2-i (\alpha +\beta )}  \nn\\
=-\frac{1}{\pi^{\frac{3}{2}}} \frac{\pi }{2 \sqrt{-i \gamma }} \int_0^\infty dx_1  \frac{\pi }{2 \sqrt{x_1^2-i (\alpha +\beta )}}   \frac{1}{\left(x_1^2-i \beta \right)}
%&=- \frac{1}{\pi^{\frac{3}{2}}} \frac{\pi }{2 \sqrt{-i \gamma }} \left[ \frac{i \pi  \arctan  \frac{\sqrt{\alpha } x_1}{\sqrt{\beta } \sqrt{x_1^2-i (\alpha +\beta )}} }{2 \sqrt{\alpha } \sqrt{\beta }}\right]_{x_1=0}^\infty\nn\\
% &=- \frac{i\sqrt{\pi}}{4  \sqrt{\alpha }\sqrt{\beta }\sqrt{-i \gamma }} \left[   \arctan  \frac{\sqrt{\alpha }  }{\sqrt{\beta } \sqrt{1-i \frac{\alpha +\beta}{x_1^2} }}\right]_{x_1=0}^\infty\nn\\
=-\frac{ \sqrt{\pi}  }{4  \sqrt{-i\alpha} \sqrt{-i\beta} \sqrt{-i\gamma}}   \left[\arctan   \frac{\sqrt{-i\alpha }  }{\sqrt{-i\beta } \sqrt{1-i \frac{\alpha +\beta}{x_1^2} }}  \right]_{x_1=0}^\infty.\label{eq:Q2}
%old: &= \frac{i\sqrt{\pi}}{4  \sqrt{\alpha }\sqrt{\beta }\sqrt{-i \gamma }} \left[   \arctan  \frac{\sqrt{\alpha } x_1}{\sqrt{\beta } \sqrt{x_1^2-i (\alpha +\beta )}}\right]_{x_1=0}^\infty\nn\\
%&=\frac{i\sqrt{\pi}}{4  e^{i\frac{\pi}{4}[2- \sgn(\alpha)-\sgn(\beta)-\sgn(\gamma)] }  \sqrt{|\alpha\beta \gamma |}}   \arctan \left(\frac{e^{i\frac{\pi}{4}[1- \sgn(\alpha)] }}{e^{i\frac{\pi}{4}[1- \sgn(\beta)] }} \sqrt{ \left|\frac{ \alpha }{ \beta  }\right|} \right)\nn\\
%&=\frac{ \sqrt{\pi}e^{i\frac{\pi}{4}[ \sgn(\alpha)+\sgn(\beta)+\sgn(\gamma)]} }{4  \sqrt{|\alpha\beta \gamma |}}   \arctan \left( e^{i\frac{\pi}{4}[- \sgn(\alpha)+  \sgn(\beta)] } \sqrt{ \left|\frac{ \alpha }{ \beta  }\right|} \right).
\end{align}
Note that in the last equality the arctan function as an antiderivative should be understood as a multivalued function not restricted to a single branch, and it must change continuously along the integration path from $x_1=0$ to $x_1=\infty$ in order to ensure that its derivative (the integrand) is finite. Thus, in addition to plugging in the principal values of the arctan function at the two limits, one must also examine carefully how its argument changes along the integration path. Whenever this argument crosses a branch cut at $(-i\infty,-i)$ or at $(i,i\infty)$, an additional term $-\pi$ or $\pi$ should be added to the result from principal values if the crossing is in a counterclockwise or clockwise sense, respectively (viewed from the origin of the complex plane). \begin{comment}
\begin{align}\label{}
&Q_2 =-\frac{i\sqrt{\pi}}{4  e^{i\frac{\pi}{4}[2- \sgn(\alpha)-\sgn(\beta)-\sgn(\gamma)] }  \sqrt{|\alpha\beta \gamma |}}   \nn\\
&\times\left[\arctan \left(\frac{e^{i\frac{\pi}{4}[1- \sgn(\alpha)] }}{e^{i\frac{\pi}{4}[1- \sgn(\beta)] }}\frac{\sqrt{|\alpha| }  }{\sqrt{|\beta |} \sqrt{1-i \frac{\alpha +\beta}{x_1^2} }} \right)\right]_{x_1=0}^\infty\nn\\
& =-\frac{ \sqrt{\pi}e^{i\frac{\pi}{4}[ \sgn(\alpha)+\sgn(\beta)+\sgn(\gamma)]} }{4  \sqrt{|\alpha\beta \gamma |}}    \nn\\
&\times\left[\arctan \left(\frac{e^{i\frac{\pi}{4}\sgn(\beta) }}{e^{i\frac{\pi}{4} \sgn(\alpha) }}\frac{\sqrt{|\alpha| }  }{\sqrt{|\beta |} \sqrt{1-i \frac{\alpha +\beta}{x_1^2} }} \right)\right]_{x_1=0}^\infty\nn\\
%&=\frac{ \sqrt{\pi}e^{i\frac{\pi}{4}[ \sgn(\alpha)+\sgn(\beta)+\sgn(\gamma)]} }{4  \sqrt{|\alpha\beta \gamma |}}   \arctan \left( e^{i\frac{\pi}{4}[- \sgn(\alpha)+  \sgn(\beta)] } \sqrt{ \left|\frac{ \alpha }{ \beta  }\right|} \right).
&=-\frac{ \sqrt{\pi}  }{4  \sqrt{-i\alpha} \sqrt{-i\beta} \sqrt{-i\gamma}}    \nn\\
&\times\left[\arctan   \frac{\sqrt{-i\alpha }  }{\sqrt{-i\beta } \sqrt{1-i \frac{\alpha +\beta}{x_1^2} }}  \right]_{x_1=0}^\infty.
\end{align}
\end{comment}

With these in mind, we analyze arguments of the arctan function appeared in \eqref{eq:Q2}. When $x_1$ changes from $x_1=0$ to $x_1=\infty$, the point  $\sqrt{|\alpha/\beta|}/\sqrt{1-i (\alpha +\beta)/x_1^2  }$ on the complex plane moves from $0$ to $|\alpha/\beta|$ along a path in the upper/lower right half plane for $\alpha+\beta$ being positive/negative, respectively, with its modulus increasing monotonically with $x_1$. The phase of $\sqrt{-i\alpha }  /\sqrt{-i\beta}$ is $-(\pi/4)[\sgn(\alpha)  -\sgn(\beta)]$, whose effect on the path is a global rotation around the origin. Below we discuss different cases:\\
1. If $|\alpha/\beta|<1$. the path will never reach a branch cut. \\
2. If $|\alpha/\beta|>1$ and $\sgn\alpha=\sgn\beta$, the additional  phase is $0$, and the path will still not reach a branch cut.\\
3. If $|\alpha/\beta|>1$ and $\sgn\alpha\ne\sgn\beta$, then if $\sgn\alpha=- \sgn\beta=1$ (this ensures $\alpha+\beta>0$), the path will reach the lower branch cut in the clockwise sense, and this produces an additional term $\pi$. On the other hand,  if $\sgn\alpha=- \sgn\beta=-1$ (this ensures $\alpha+\beta<0$), the path will reach the upper branch cut in the counterclockwise sense, but does not cross it. We follow the conventional definition to take the value at a point exactly on a branch cut to be continuous when approached from the counterclockwise direction. So no additional term is added.

In sum, only in the case of  $|\alpha/\beta|>1$ and $\sgn\alpha=-\sgn\beta=1$  will there be an additional term $\pi$. This is equivalent to the requirement that  $\beta<0$ and $\alpha+\beta>0$.
\begin{comment}
Thus, we arrive at:
\begin{align}\label{}
&\left[\arctan \left(\frac{e^{i\frac{\pi}{4}\sgn(\beta) }}{e^{i\frac{\pi}{4} \sgn(\alpha) }}\frac{\sqrt{|\alpha| }  }{\sqrt{|\beta |} \sqrt{1-i \frac{\alpha +\beta}{x_1^2} }} \right)\right]_{x_1=0}^\infty\nn\\
&= \theta(-\beta)\theta(\alpha+\beta)\pi+ \operatorname{Arctan} \left(\frac{e^{i\frac{\pi}{4}\sgn(\beta) }}{e^{i\frac{\pi}{4} \sgn(\alpha) }}\frac{\sqrt{|\alpha| }  }{\sqrt{|\beta |} } \right),
\end{align}
where ``Arctan'' is used to denote the principal value of arctan.
\end{comment}
Therefore, we obtain the final result of $Q_2$:
\begin{align}\label{}
&Q_2  =-\frac{ \sqrt{\pi} }{4  \sqrt{-i\alpha} \sqrt{-i\beta}\sqrt{-i\gamma }}
 \left[ \theta(-\beta)\theta(\alpha+\beta)\pi+ \operatorname{Arctan}  \frac{\sqrt{-i\alpha }  }{\sqrt{-i\beta }  }\right],
\end{align}
where ``Arctan'' is used to denote the principal value of arctan.

Since the integrand of $Q_3$ can be obtained from that of $Q_2$ by exchanging $x_1$ with $x_2$ and $\beta$ with $\gamma$, so $Q_3$ can be obtained from $Q_2$ simply by switching $\beta$ and $\gamma$ in the result of $Q_2$:
\begin{align}\label{}
&Q_3  %=-\frac{ \sqrt{\pi}  }{4  \sqrt{-i\alpha} \sqrt{-i\beta} \sqrt{-i\gamma}}    \nn\\
%&\times\left[\arctan  \frac{\sqrt{-i\alpha }  }{\sqrt{-i\gamma } \sqrt{1-i \frac{\alpha +\gamma}{x_2^2} }}  \right]_{x_2=0}^\infty\nn\\
=-\frac{ \sqrt{\pi} }{4  \sqrt{-i\alpha} \sqrt{-i\beta}\sqrt{-i\gamma }}
  \left[ \theta(-\gamma)\theta(\alpha+\gamma)\pi+ \operatorname{Arctan}  \frac{\sqrt{-i\alpha }  }{\sqrt{-i\gamma }  }\right],
\end{align}

Finally we evaluate $Q_4$, which is the most complicated. Performing first the integration on $x$ gives:
\begin{align}\label{}
&Q_4%=\frac{1}{\pi^{\frac{3}{2}}} \int_0^\infty dx\int_0^\infty dx_1 \int_0^\infty dx_2  \frac{1}{\left(x_1^2-i \beta \right) \left(x_2^2-i \gamma \right)}\nn\\
%&\times\frac{1}{x^2+x_1^2+x_2^2-i (\alpha +\beta +\gamma )} \nn\\
=\frac{1}{\pi^{\frac{3}{2}}}  \int_0^\infty dx_1 \int_0^\infty dx_2 \frac{1}{\left(x_1^2-i \beta \right) \left(x_2^2-i \gamma \right)}
\frac{\pi}{2\sqrt{ x_1^2+x_2^2-i (\alpha +\beta +\gamma )}}.
\end{align}
Introduce polar coordinates $x_1=r\cos\varphi$ and $x_2=r\sin\varphi$, $Q_4$ is rewritten as
\begin{align}\label{eq:Q4-1}
&Q_4%=\frac{1}{\pi^{\frac{3}{2}}} \int_0^\infty dx\int_0^\infty dx_1 \int_0^\infty dx_2  \frac{1}{\left(x_1^2-i \beta \right) \left(x_2^2-i \gamma \right)}\nn\\
%&\times\frac{1}{x^2+x_1^2+x_2^2-i (\alpha +\beta +\gamma )} \nn\\
=\frac{1}{\pi^{\frac{3}{2}}}  \int_0^\infty dr r  \frac{\pi}{2\sqrt{ r^2-i (\alpha +\beta +\gamma )}} I_\varphi,
\end{align}
where we denoted the angular integral  as
\begin{align}\label{}
I_\varphi\equiv\int_0^{\frac{\pi}{2}} d\varphi \frac{1}{\left(r^2\cos^2\varphi-i \beta \right) \left(r^2\sin^2\varphi-i \gamma \right)}.
\end{align}
$I_\varphi$ can be performed by contour integration. Using symmetry to extend the integral region and changing variable to $\theta=2\varphi$, we get
\begin{align}\label{}
&I_\varphi%=\int_0^{\frac{\pi}{2}} d\varphi \frac{1}{\left(r^2\cos^2\varphi-i \beta \right) \left(r^2\sin^2\varphi-i \gamma \right)},\nn\\
%&=\frac{1}{2}\int_0^{\pi} d\varphi \frac{1}{\left(r^2\cos^2\varphi-i \beta \right) \left(r^2\sin^2\varphi-i \gamma \right)}\nn\\
=\frac{1}{4}\int_0^{2\pi}  d\theta \frac{1}{\left[\frac{r^2}{2}(1+\cos\theta)-i \beta \right] \left[\frac{r^2}{2}(1-\cos\theta)-i \gamma \right]}.
\end{align}
A transformation $z=e^{i\theta}$ then recasts the integral as one along a unit circle of an complex variable $z$:
\begin{align}\label{}
I_\varphi=\frac{1}{4}\oint_{|z|=1} \frac{dz}{iz}  \frac{1}{\left[\frac{r^2}{2}(1+\frac{z^2+1}{2z})-i \beta \right] \left[\frac{r^2}{2}(1-\frac{z^2+1}{2z})-i \gamma \right]}.
\end{align}
Application of the residue theorem then gives:
\begin{align}\label{}
&I_\varphi=\frac{\pi}{2}\sum_{|z|<1} \operatorname{res} \frac{1}{z\left[\frac{r^2}{2}(1+\frac{z^2+1}{2z})-i \beta \right] \left[\frac{r^2}{2}(1-\frac{z^2+1}{2z})-i \gamma \right]},
%&=\frac{\pi}{2}\sum_{|z|<1} \operatorname{res}\left\{\frac{4z}{\left[\frac{r^2}{2}(z^2+2z+1)-i2 \beta z\right] \left[\frac{r^2}{2}(-z^2+2z-1)-i 2\gamma z\right]}\right\}\nn\\
%&=-\frac{8\pi}{r^4}\sum_{|z|<1} \operatorname{res}\left\{\frac{z}{\left[  z^2+2(1-i \frac{2 \beta}{r^2})z+1   \right] \left[  z^2-2(1-i \frac{ 2\gamma}{r^2})z+1   \right]}\right\}.
\end{align}
where the sum is over all singular points inside the unit circle. The  singular points are located at the roots of two equations:
\begin{align}\label{}
& z^2+2(1-i \frac{ 2\beta}{r^2})z+1  =0,\\
& z^2-2(1-i \frac{ 2\gamma}{r^2})z+1    =0,
\end{align}
which are
\begin{align}\label{}
& z_{1,2}=-1+i \frac{2 \beta}{r^2} \pm\sqrt{\left(1-i \frac{2 \beta}{r^2}\right)^2-1},\\
& z_{3,4}=1-i \frac{2 \gamma}{r^2} \pm\sqrt{\left(1-i \frac{ 2\gamma}{r^2}\right)^2-1}.
\end{align}
Only the two roots $z_1$ and $z_4$ are in the unit circle and contribute to the sum of residues. Calculating out these two residues, we obtain
\begin{align}\label{}
&I_\varphi%=%-\frac{8\pi}{r^4}\nn\\
%&\times\left\{\left.\frac{z}{2 (  z+ 1-i \frac{2 \beta}{r^2} +\sqrt{\left(1-i \frac{2 \beta}{r^2}\right)^2-1} ) \left[  z^2-2(1-i \frac{2 \gamma}{r^2})z+1   \right]}\right|_{z=z_1}\right.\nn\\
%&+\left.\left.\frac{z}{\left[  z^2+2(1-i \frac{2 \beta}{r^2})z+1   \right] 2 (  z-1+i \frac{ 2\gamma}{r^2} -\sqrt{\left(1-i \frac{2\gamma}{r^2}\right)^2-1} )}\right|_{z=z_4}\right\}\nn\\
%&= -\frac{8\pi}{r^4}\times\frac{-r^4}{16 \left[r^2-i (\beta +\gamma )\right]}\left[\frac{1}{\sqrt{-\beta  \left(\beta +i r^2\right)}}+\frac{1}{\sqrt{-\gamma  \left(\gamma +i r^2\right)}}\right]\nn\\
=  \frac{\pi}{2 \left[r^2-i (\beta +\gamma )\right]}\left[\frac{1}{\sqrt{-\beta  \left(\beta +i r^2\right)}}+\frac{1}{\sqrt{-\gamma  \left(\gamma +i r^2\right)}}\right].
\end{align}
Plugging into \eqref{eq:Q4-1} and changing variable to $y=r^2$, we get $Q_4$  expressed as a single integral:
\begin{align}\label{eq:Q4}
&Q_4%=\frac{1}{\pi^{\frac{3}{2}}}  \int_0^\infty dr r  \frac{\pi}{2\sqrt{ r^2-i (\alpha +\beta +\gamma )}} \nn\\
%&\times\frac{\pi}{2 \left[r^2-i (\beta +\gamma )\right]}\left[\frac{1}{\sqrt{-\beta  \left(\beta +i r^2\right)}}+\frac{1}{\sqrt{-\gamma  \left(\gamma +i r^2\right)}}\right]\nn\\
%&=\frac{\sqrt{\pi}}{4}  \int_0^\infty dr r  \frac{1}{ \sqrt{ r^2-i (\alpha +\beta +\gamma )}} \nn\\
%&\times\frac{1}{  r^2-i (\beta +\gamma )}\left[\frac{1}{\sqrt{-\beta  \left(\beta +i r^2\right)}}+\frac{1}{\sqrt{-\gamma  \left(\gamma +i r^2\right)}}\right]\nn\\
=\frac{\sqrt{\pi}}{8}  \int_0^\infty dy \frac{1}{ \sqrt{ y-i (\alpha +\beta +\gamma )}}
 \frac{1}{  y-i (\beta +\gamma ) }\left[\frac{1}{\sqrt{-\beta  \left(\beta +iy\right)}}+\frac{1}{\sqrt{-\gamma  \left(\gamma +i y\right)}}\right].
\end{align}
The two terms in the integrand are symmetric in $\beta$ and $\gamma$. For the first term, direct integration gives:
\begin{align}\label{}
 &I_y\equiv\int_0^\infty dy \frac{1}{ \sqrt{ y-i (\alpha +\beta +\gamma )}}  \frac{1}{   y-i (\beta +\gamma ) } \frac{1}{\sqrt{-\beta  \left(\beta +iy\right)}}
% &=\int_0^\infty dy \frac{1}{ \sqrt{ y-i (\alpha +\beta +\gamma )}}  \frac{1}{   y-i (\beta +\gamma ) } \frac{1}{\sqrt{-i\beta}\sqrt{y-i\beta}}\nn\\
 % old: &=\left[\frac{2 \sqrt{\beta +i y} \arctan  \frac{\sqrt{-i \alpha } \sqrt{\beta +i y}}{\sqrt{\gamma } \sqrt{y-i (\alpha +\beta +\gamma )}} }{\sqrt{-i \alpha } \sqrt{\gamma } \sqrt{-i \beta  (y-i \beta )}}\right]_{y=0}^\infty.
%&=\left[\frac{2 \alpha  \sqrt{-i \gamma } \arctan \frac{\sqrt{-i \alpha } \sqrt{y-i \beta }}{\sqrt{-i \gamma } \sqrt{y-i (\alpha +\beta +\gamma )}} }{(-i \alpha )^{3/2} \sqrt{-i \beta } \gamma }\right]_{y=0}^\infty\nn\\
=\frac{2 }{  \sqrt{-i \alpha } \sqrt{-i \beta } \sqrt{-i \gamma } }\left[\arctan \frac{\sqrt{-i \alpha } \sqrt{y-i \beta }}{\sqrt{-i \gamma } \sqrt{y-i (\alpha +\beta +\gamma )}}   \right]_{y=0}^\infty,
\end{align}
where we used the fact that since $\beta$ is real and $y$ is positive, $ \sqrt{-\beta  (\beta +iy)} =\sqrt{-i\beta}\sqrt{y-i\beta}  $ always holds. The second term in \eqref{eq:Q4} can be obtained from $I_y$ by switching $\beta$ and $\gamma$. We thus obtain $Q_4$:
\begin{align}\label{}
&Q_4=\frac{\sqrt{\pi}}{8} [I_y+(\beta\leftrightarrow\gamma)]
=\frac{\sqrt{\pi}}{ 4 \sqrt{-i \alpha } \sqrt{-i \beta } \sqrt{-i \gamma } }\left[\arctan \frac{\sqrt{-i \alpha } \sqrt{y-i \beta }}{\sqrt{-i \gamma } \sqrt{y-i (\alpha +\beta +\gamma )}}
 + \arctan \frac{\sqrt{-i \alpha } \sqrt{y-i \gamma}}{\sqrt{-i \beta  } \sqrt{y-i (\alpha +\beta +\gamma )}}  \right]_{y=0}^\infty.
\end{align}
Again, the arguments of the arctan functions need careful treatment. %After careful treatment of the square roots, we get:
Instead of treating directly the arctan function in $Q_4$, it turns out to be more convenient to first combine the results of $Q_2$, $Q_3$ and $Q_4$ for in terms of the antiderivatives:
\begin{align}\label{}
&Q_2+Q_3+Q_4=\frac{\sqrt{\pi}}{ 4 \sqrt{-i \alpha } \sqrt{-i \beta } \sqrt{-i \gamma } }\nn\\
&\times\left[-\arctan   \frac{\sqrt{-i\alpha }  }{\sqrt{-i\beta } \sqrt{1-i \frac{\alpha +\beta}{y} }} -\arctan   \frac{\sqrt{-i\alpha }  }{\sqrt{-i\gamma } \sqrt{1-i \frac{\alpha +\gamma}{y} }} \right.+\arctan \frac{\sqrt{-i \alpha } \sqrt{y-i \beta }}{\sqrt{-i \gamma } \sqrt{y-i (\alpha +\beta +\gamma )}}   \nn\\
&\left.+ \arctan \frac{\sqrt{-i \alpha } \sqrt{y-i \gamma}}{\sqrt{-i \beta  } \sqrt{y-i (\alpha +\beta +\gamma )}}  \right]_{y=0}^\infty
\end{align}
where we have changed the dummy integration variables in $Q_2$ and $Q_3$ from $x_1^2$ and $x_2^2$ to $y$. The second/third term can be obtained from the first/fourth term respectively by exchanging $\beta$ and $\gamma$, so it suffices to consider the first and fourth terms. We also reverse the two limits in the fourth term, so
\begin{align}\label{eq:Q2Q3Q4}
&Q_2+Q_3+Q_4=-\frac{\sqrt{\pi}}{ 4 \sqrt{-i \alpha } \sqrt{-i \beta } \sqrt{-i \gamma } }\nn\\
&\times\left\{\left[\arctan   \frac{\sqrt{-i\alpha }  }{\sqrt{-i\beta } \sqrt{1-i \frac{\alpha +\beta}{y} }} \right]_{y=0}^\infty
 +\left[  \arctan \frac{\sqrt{-i \alpha } \sqrt{y-i \gamma}}{\sqrt{-i \beta  } \sqrt{y-i (\alpha +\beta +\gamma )}}  \right]_{y=\infty}^0\right\}+(\beta\leftrightarrow\gamma).
\end{align}
We now need to analyze at which condition will the path make a finite contribution to the result from principal values. In \eqref{eq:Q2Q3Q4} the argument of the first arctan function moves from $0$ at $y=0$ to $\sqrt{-i\alpha}/\sqrt{-i\beta}$ at $y=\infty$, and the argument of the second arctan function moves from $\sqrt{-i\alpha}/\sqrt{-i\beta}$ at $y=\infty$ to $\sqrt{-i\alpha}\sqrt{-i\gamma}/[\sqrt{-i\beta}\sqrt{-i(\alpha+\beta+\gamma)}]$ at $y=0$. So the two paths are connected to a single one. Plotting at different choices of $\alpha$, $\beta$ and $\gamma$ shows that such a connected path always lies in a single quadrant of the complex plane, so it can never cross a branch cut. But it can end exactly on a branch cut. This can only happen when the final point's modulus satisfies $|(\alpha \gamma)/[\beta(\alpha+\beta+\gamma)]|>1$ and its phase satisfies $\arg[\sqrt{-i\alpha}\sqrt{-i\gamma}/(\sqrt{-i\beta}\sqrt{-i(\alpha+\beta+\gamma)})]=\pi[-\sgn(\alpha)+\sgn(\beta)-\sgn(\gamma)+\sgn(\alpha+\beta+\gamma)]/4 =\pm \pi/2$, where the upper/lower signs correspond to the upper and lower branch cuts, respectively. Further, according to the definition of values on a branch cut, in order to produce a finite contribution to the result, the path must approach the branch cut in the clockwise sense. This can only takes place when the path from $0$ to $\sqrt{-i\alpha}/\sqrt{-i\beta}$ is in the second/fourth quadrant if the path ends at the upper/lower branch cut, respectively. Looking at the value of the argument at $y=0^+$ gives the condition $\pi[-\sgn(\alpha)+\sgn(\beta)+\sgn(\alpha+\beta)]/4=3\pi/4$ or $-\pi/4$. Setting it to $3\pi/4$ gives $\alpha<0$ and $\alpha+\beta>0$, and setting it to $-\pi/4$ gives $\alpha<0$ and $\beta<0$, or $\beta<0$ and $\alpha+\beta>0$. Thus, there are two cases for the signs. Case 1 is (recall that $\alpha+\beta+\gamma>0$)
\begin{align}\label{eq:sign-cond1}
&-\sgn(\alpha)+\sgn(\beta)-\sgn(\gamma)=1,\nn\\
&-\sgn(\alpha)+\sgn(\beta)+\sgn(\alpha+\beta)=3,
\end{align}
Case 2 is
\begin{align}\label{eq:sign-cond2}
&-\sgn(\alpha)+\sgn(\beta)-\sgn(\gamma)=-3,\nn\\
&-\sgn(\alpha)+\sgn(\beta)+\sgn(\alpha+\beta)=-1.
\end{align}
Cases 1 gives: $\alpha<0$, $\beta>0$, $\gamma>0$, and $\alpha+\beta>0$. Cases 2 gives: $\alpha>0$, $\beta<0$, $\gamma>0$, and $\alpha+\beta>0$. But note that for case 1, one always has $|\alpha|<|\beta|$ and $|\gamma|<|\alpha+\beta+\gamma|$, so $|(\alpha \gamma)/[\beta(\alpha+\beta+\gamma)]|>1$ can never be satisfied, so case 1 is not possible. For case 2, one has $|\alpha|>|\beta|$ and $|\gamma|<|\alpha+\beta+\gamma|$, and $|(\alpha \gamma)/[\beta(\alpha+\beta+\gamma)]|>1$ reduces to $-\alpha\gamma/[\beta(\alpha+\beta+\gamma)]>1$, or $(\alpha+\beta)(\beta+\gamma)>0$. This together with $\alpha+\beta>0$ gives $\beta+\gamma>0$. In sum, only in the case of  $\beta<0$, $\alpha+\beta>0$ and $\beta+\gamma>0$  will there be an additional term $\pi$ added to the result of the first two arctan functions in \eqref{eq:Q2Q3Q4}. Thus,
\begin{align}\label{eq:first-two-arctan}
&\left[\arctan   \frac{\sqrt{-i\alpha }  }{\sqrt{-i\beta } \sqrt{1-i \frac{\alpha +\beta}{y} }} \right]_{y=0}^\infty
+\left[  \arctan \frac{\sqrt{-i \alpha } \sqrt{y-i \gamma}}{\sqrt{-i \beta  } \sqrt{y-i (\alpha +\beta +\gamma )}}  \right]_{y=\infty}\nn\\
&= \theta(-\beta)\theta(\alpha+\beta)\theta(\beta+\gamma)\pi
+ \operatorname{Arctan} \frac{\sqrt{-i \alpha } \sqrt{ -i \gamma}}{\sqrt{-i \beta  } \sqrt{ -i (\alpha +\beta +\gamma )}} .
\end{align}
The term with step functions affects only those cases at which the argument of the Arctan function lies on the lower branch cut and satisfy $\beta<0$, $\alpha+\beta>0$ and $\beta+\gamma>0$. Note an Arctan function has a real part within $[-\pi/2, \pi/2]$, and when its argument is on the lower branch cut its real part is $-\pi/2$. Thus, the result of \eqref{eq:first-two-arctan} still has a real part within $[-\pi/2, \pi/2]$. The conditions for which a $\pi$ needs to be added to the ``$(\beta\leftrightarrow\gamma)$'' terms can be obtained directly by switching $\beta$ with $\gamma$. Finally we get:
\begin{align}\label{eq:Q2Q3Q4-2}
&Q_2+Q_3+Q_4=-\frac{\sqrt{\pi}}{ 4 \sqrt{-i \alpha } \sqrt{-i \beta } \sqrt{-i \gamma } }
\left[\theta(-\beta)\theta(\alpha+\beta)\theta(\beta+\gamma)\pi
+ \operatorname{Arctan} \frac{\sqrt{-i \alpha } \sqrt{ -i \gamma}}{\sqrt{-i \beta  } \sqrt{ -i (\alpha +\beta +\gamma )}} \right] +(\beta\leftrightarrow\gamma). %\nn\\
%&=-\frac{\sqrt{\pi}}{ 4 \sqrt{-i \alpha } \sqrt{-i \beta } \sqrt{-i \gamma } }\nn\\
%&\times  \left\{[\theta(-\beta)\theta(\alpha+\beta)+\theta(-\gamma)\theta(\alpha+\gamma)]\theta(\beta+\gamma)\pi\right.\nn\\
%&\left.+ \operatorname{Arctan} \frac{\sqrt{-i \alpha } \sqrt{ -i \gamma}}{\sqrt{-i \beta  } \sqrt{ -i (\alpha +\beta +\gamma )}}+\operatorname{Arctan} \frac{\sqrt{-i \alpha } \sqrt{ -i \beta}}{\sqrt{-i \gamma  } \sqrt{ -i (\alpha +\beta +\gamma )}} \right\} .
\end{align}
The two arctan functions can be combined by using
\begin{align}\label{}
&\Arctan(z_1)+\Arctan(z_2)=\Arctan\frac{z_1+z_2}{1-z_1 z_2}+ M \pi,
\end{align}
where the additional $ M \pi$ term with $M $ an integer is needed to connect values at different branches. %when $\Arctan(z_1)+\Arctan(z_2)$ does not lie within $[\pi/2,\pi/2]$... %which is needed to connect the functions at different branches.
Here we have $z_1= \sqrt{-i \alpha } \sqrt{-i \beta } /[ \sqrt{-i \gamma }\sqrt{-i(\alpha+\beta+\gamma)}]$ and $z_2= \sqrt{-i \alpha } \sqrt{-i \gamma } /[ \sqrt{-i \beta }\sqrt{-i(\alpha+\beta+\gamma)}]$, and $(z_1+z_2)/(1-z_1 z_2)$ can be reduced to a simple form:
\begin{align}\label{}
&\frac{(z_1+z_2)}{1-z_1 z_2}%=\frac{   \frac{\sqrt{-i \alpha } \sqrt{-i \beta } }{ \sqrt{-i \gamma }\sqrt{-i(\alpha+\beta+\gamma)} }  +  \frac{\sqrt{-i \alpha } \sqrt{-i \gamma } }{ \sqrt{-i \beta }\sqrt{-i(\alpha+\beta+\gamma)} }  }{1-     \frac{\sqrt{-i \alpha } \sqrt{-i \beta } }{ \sqrt{-i \gamma }\sqrt{-i(\alpha+\beta+\gamma)} }      \frac{\sqrt{-i \alpha } \sqrt{-i \gamma } }{ \sqrt{-i \beta }\sqrt{-i(\alpha+\beta+\gamma)} }   }\nn\\
%&=\frac{\frac{ \sqrt{-i \alpha } }{\sqrt{-i(\alpha+\beta+\gamma)}} \left(\frac{  \sqrt{-i \beta } }{ \sqrt{-i \gamma }   }  +  \frac{  \sqrt{-i \gamma } }{ \sqrt{-i \beta } }\right)  } {1-     \frac{ -i \alpha }{-i(\alpha+\beta+\gamma)}    }\nn\\
%& =\frac{\frac{ \sqrt{-i \alpha } \sqrt{-i(\alpha+\beta+\gamma)}}{ \sqrt{-i \beta }  \sqrt{-i \gamma }  } (-i \beta -i \gamma) } {-i(\alpha+\beta+\gamma)     +i \alpha   }\nn\\
 = \frac{ \sqrt{-i \alpha } \sqrt{-i(\alpha+\beta+\gamma)}}{ \sqrt{-i \beta }  \sqrt{-i \gamma }  }   .
\end{align}
The integer $M$ can be determined by calculating out numerical values at different choices $\alpha$, $\beta$ and $\gamma$. It turns out that $M =1$ when $\alpha+\beta>0$, $\alpha+\gamma>0$ and $\beta+\gamma<0$ or when $\alpha+\beta<0$ and $\alpha+\gamma<0$ are satisfied, and $M =0$ otherwise. Thus,
\begin{align}\label{}
M=\theta(\alpha+\beta)\theta(\alpha+\gamma)\theta(-\beta-\gamma)+\theta (- \alpha-\beta) \theta (- \alpha-\gamma).
\end{align}
This contribution from $M\pi$ can be combined with the term with step functions in \eqref{eq:Q2Q3Q4-2}, and we arrive at
\begin{align}\label{eq:Q2Q3Q4-3}
&Q_2+Q_3+Q_4=-\frac{\sqrt{\pi}}{ 4 \sqrt{-i \alpha } \sqrt{-i \beta } \sqrt{-i \gamma } }\nn\\
&\times  \left\{\{\theta[(\alpha+\beta)(\alpha+\gamma)]-\theta (\beta ) \theta (\gamma ) \theta (\alpha+\beta ) \theta (\alpha+\gamma )\} \pi
+ \Arctan  \frac{ \sqrt{-i \alpha } \sqrt{-i(\alpha+\beta+\gamma)}}{ \sqrt{-i \beta }  \sqrt{-i \gamma }  } \right\} ,
\end{align}
which can further be  written as
\begin{align}\label{eq:Q2Q3Q4-4}
&Q_2+Q_3+Q_4=-\frac{\sqrt{\pi}}{ 4 \sqrt{-i \alpha } \sqrt{-i \beta } \sqrt{-i \gamma } }
  \left\{\left[\frac{1}{2} -\theta (-\alpha)\theta(\alpha+\beta)\theta (\alpha+\gamma) \right] \pi
  - \Arctan  \frac{ \sqrt{-i \beta }  \sqrt{-i \gamma }  }{ \sqrt{-i \alpha } \sqrt{-i(\alpha+\beta+\gamma)}} \right\} .
\end{align}
Combining with $Q_1$, finally we obtain $Q$:
\begin{align}\label{Q-result-general}
&Q=\frac{\sqrt{\pi}}{ 4 \sqrt{-i \alpha } \sqrt{-i \beta } \sqrt{-i \gamma } }  \left[ \theta (-\alpha)\theta(\alpha+\beta)\theta (\alpha+\gamma)   \pi
+ \Arctan  \frac{ \sqrt{-i \beta }  \sqrt{-i \gamma }  }{ \sqrt{-i \alpha } \sqrt{-i(\alpha+\beta+\gamma)}} \right].
\end{align}
\begin{comment}
\begin{align}\label{Q-result-general}
&Q =   \frac{\sqrt{\pi}  }{4 \sqrt{-i \alpha }    \sqrt{-i \beta }  \sqrt{-i \gamma }} \nn\\
 & \times\left[\frac{\pi}{2}- \arctan  \frac{ \sqrt{-i \alpha } \sqrt{-i(\alpha+\beta+\gamma)}}{ \sqrt{-i \beta }  \sqrt{-i \gamma }  }\right].
\end{align}
\end{comment}
Writing $\alpha$, $\beta$ and $\gamma$ back to the slopes, we get \eqref{Q-result-off-res}.

We checked  at different sets of values of $\alpha$, $\beta$ and $\gamma$ that the expression \eqref{Q-result-general} agrees with results of numerical integration of \eqref{eq:Q}. But the Arctan function in \eqref{Q-result-general} diverges at special choices of $\alpha$, $\beta$ and $\gamma$ when its argument happens to be at the branch points $\pm i$, and  \eqref{Q-result-general} does not apply in this special case. \begin{comment}
This happens when two conditions,
\begin{equation}\label{eq:arctan-phase}
\frac{e^{i\frac{\pi}{4}[ \sgn(\beta)+\sgn(\gamma)]}}{e^{i\frac{\pi}{4}[ \sgn(\alpha)+\sgn(\alpha+\beta+\gamma)]}}=\pm i
\end{equation}
and
\begin{equation}\label{eq:arctan-modulus}
\left|\frac{ \beta \gamma}{\alpha(\alpha+\beta+\gamma) }\right|=1,
\end{equation}
are simultaneously satisfied. Note that at $  \beta \gamma/[\alpha(\alpha+\beta+\gamma)] =1$ the left-hand side of \eqref{eq:arctan-phase} can only take $\pm 1$, so the second condition \eqref{eq:arctan-modulus} reduces to $\beta \gamma/[\alpha(\alpha+\beta+\gamma)]  =-1$, or
\begin{equation}\label{}
(\alpha+\beta)(\alpha+\gamma)=0.
\end{equation}
At either $\alpha+\beta=0$ or at $\alpha+\gamma=0$ the left-hand side of \eqref{eq:arctan-phase} equals $-i\sgn(\alpha)$, so \eqref{eq:arctan-phase} is also satisfied. Therefore,
\end{comment}
This takes place when
\begin{align}\label{}
\left[ \frac{ \sqrt{-i \beta }  \sqrt{-i \gamma }  }{ \sqrt{-i \alpha } \sqrt{-i(\alpha+\beta+\gamma)}} \right]^2=\frac{\beta \gamma}{\alpha(\alpha+\beta+\gamma)}=-1,
\end{align}
or
\begin{align}\label{}
(\alpha+\beta)(\alpha+\gamma)=0.
\end{align}
Therefore, \eqref{Q-result-general} is not applicable whenever any of the two conditions
\begin{align}\label{}
&\alpha+\beta=0,\\
&\alpha+\gamma=0,
\end{align}
are satisfied. Writing them in terms of the slopes, we arrive at the resonance condition discussed in the main text.

\section*{Appendix C: $4$-state LZ model with all-to-all couplings}

\setcounter{figure}{0}
\setcounter{equation}{0}
\renewcommand{\theequation}{C\arabic{equation}}
\renewcommand\thefigure{C\arabic{figure}}

In this appendix, we present perturbative results for the most general one-crossing $4$-state LZ model, with a Hamiltonian:
\begin{align}\label{}
& H= \left( \begin{array}{cccc}
 b_1 t   &  A_{12}    &  A_{13} &  A_{14}\\
A_{12} &  b_2 t &  A_{23}  & A_{24}  \\
 A_{13} & A_{23} & b_3 t & A_{34} \\
 A_{14} & A_{24} & A_{34} & b_4 t
\end{array} \right),
\label{eq:Hal-4}
\end{align}
where $b_1>b_2>b_3>b_4$.  We obtain:
\begin{align}
&P_{12}= 2  \lambda_{12}^2-2\lambda_{12}(\lambda_{13}\lambda_{23} +\lambda_{14}\lambda_{24})
+(\lambda_{13}\lambda_{23} +\lambda_{14}\lambda_{24})^2-\lambda_{12}^2[3(\lambda_{13}^2+\lambda_{14}^2)+ \lambda_{23}^2+\lambda_{24}^2
+2\lambda_{12}^2]+O(g^5) ,\\
&P_{13}= 2  \lambda_{13}^2+2\lambda_{13}(\lambda_{12}\lambda_{23} -\lambda_{14}\lambda_{34})
+\lambda_{12}^2\lambda_{23}^2 +\lambda_{14}^2\lambda_{34}^2 -\lambda_{13}^2(\lambda_{12}^2+3\lambda_{14}^2+ \lambda_{23}^2+\lambda_{34}^2+2\lambda_{13}^2) \nn\\
&-2\lambda_{13}(2\lambda_{14}\lambda_{24}\lambda_{23} +\lambda_{12}\lambda_{24}\lambda_{34}) +O(g^5) ,\\
&P_{14}= 2  \lambda_{14}^2+2\lambda_{14}(\lambda_{12}\lambda_{24} +\lambda_{13}\lambda_{34})
+(\lambda_{12}\lambda_{24} +\lambda_{13}\lambda_{34})^2 -\lambda_{14}^2(\lambda_{12}^2+\lambda_{13}^2+ \lambda_{24}^2+\lambda_{34}^2+2\lambda_{14}^2)  \nn\\
&+2 \lambda_{14} \lambda_{13}\lambda_{23}\lambda_{24} +O(g^5) ,\\
&P_{23}= 2  \lambda_{23}^2-2\lambda_{23}(\lambda_{12}\lambda_{13} +\lambda_{24}\lambda_{34})+(\lambda_{12}\lambda_{13} +\lambda_{24}\lambda_{34})^2 -\lambda_{23}^2(\lambda_{12}^2+3\lambda_{24}^2% not \lambda_{14}^2 here!
+ \lambda_{34}^2+3\lambda_{13}^2 +2\lambda_{23}^2)\nn\\
&-2 \lambda_{23}\lambda_{24}\lambda_{14}\lambda_{13} +O(g^5) ,\\
&P_{11}= 1-2( \lambda_{12}^2+\lambda_{13}^2 +\lambda_{14}^2)+ 2 (\lambda_{12}^2+\lambda_{13}^2 + \lambda_{14}^2)^2+O(g^6) ,\label{eq:4-state-P11}\\
&P_{22}=1-2( \lambda_{12}^2+\lambda_{23}^2 +\lambda_{24}^2)
+2 \lambda_{12}^4+2 \lambda_{12}^2 \left(\lambda_{13}^2+\lambda_{14}^2+\lambda_{23}^2+\lambda_{24}^2\right)+\lambda_{23}^2 \left(2 \lambda_{13}^2+3 \lambda_{24}^2+2\lambda_{23}^2\right)\nn\\
&+\lambda_{24}^2 \left(2 \lambda_{14}^2+\lambda_{23}^2\right)
+2 \lambda_{24}^4+4 \lambda_{13} \lambda_{14}\lambda_{23}\lambda_{24}+O(g^6).
\end{align}
Again, other $P_{jk}$ with $j\le k$ are connected to these explicitly written out ones  by switches of indices $1\leftrightarrow 4$ and $2\leftrightarrow 3$ everywhere, and $P_{jk}$ with $j>k$ can be obtained by changing the signs of the $3$rd order terms in  $P_{kj}$. We do not perform numerical checks for this model, but note that $P_{11}$ in \eqref{eq:4-state-P11} agrees with the exact result from the BE formula.


\begin{thebibliography}{}
\bibitem{note-phase} When there are more than one such paths, one also needs to appropriately take into account interference of them due to their phases.
\bibitem{note-diabatic}
We make precise what is the meaning of the diabatic limit, or ``small $g$''. For one-crossing MLZ models discussed later, it means that all the LZ parameters $(g A_{jk})^2/|b_j-b_k|$ are much smaller than $1$, so that $g$ is much smaller than a value $g_c\sim \min(\sqrt{|b_j-b_k|}/|A_{jk}|)$. For the more general model Eq.~\eqref{eq:Hamiltonian}, $|b_j-b_k|$ should be replaced by the values of $|\partial (\varepsilon_j-\varepsilon_k)/\partial t|$ at the crossing of diabatic levels. If we take both the couplings $A_{jk}$ and the slope differences $|b_j-b_k|$ to be of the order of $1$ (recall that in our units system they are all dimensionless), then small $g$ simply means $g\ll 1$.
\end{thebibliography}
\end{document}